\documentclass[twocolumn]{aastex631}

\shorttitle{NSC in nearby galaxies}
\shortauthors{Ashok et al.}
\graphicspath{{./}{figures/}}
\pdfoutput=1
\graphicspath{{./}}
\usepackage{color}
\usepackage{amsmath}
\usepackage{array}
\usepackage{xfrac}
\usepackage[bottom]{footmisc}
\usepackage{lipsum}
\usepackage{tablefootnote}
\usepackage{textcomp}
\usepackage{booktabs}
\usepackage{listings}
\usepackage{graphicx}
\usepackage{float}

\newcommand{\Msun}{\ensuremath{M_{\sun}}}
\newcommand{\Mstar}{\ensuremath{M_{\star}}}
\newcommand{\Mnsc}{\ensuremath{M_{\mathrm{NSC}}}}

\definecolor{codegreen}{rgb}{0,0.6,0}
\definecolor{codegray}{rgb}{0.5,0.5,0.5}
\definecolor{codepurple}{rgb}{0.58,0,0.82}
\definecolor{backcolour}{rgb}{0.95,0.95,0.92}

\begin{document}

\title{Composite Bulges -- III. A Study of Nuclear Star Clusters in Nearby Spiral Galaxies}

\author[0000-0003-1143-8502]{Aishwarya Ashok}
\affiliation{Department of Physics and Astronomy, University of Utah, Salt Lake City, UT 84112, USA}

\author[0000-0003-0248-5470]{Anil Seth}
\affiliation{Department of Physics and Astronomy, University of Utah, Salt Lake City, UT 84112, USA}

\author[0000-0003-4588-9555]{Peter Erwin}
\affiliation{Max-Planck-Insitut f\"{u}r extraterrestrische Physik, Giessenbachstrasse, D85748 Garching, Germany}
\affiliation{Universit\"{a}ts-Sternwarte M\"{u}nchen, Scheinerstrasse 1, D-81679 M\"{u}nchen, Germany}

\author{Victor P. Debattista}
\affiliation{Jeremiah Horrocks Institute, University of Central Lancashire, Preston, PR1 2HE, UK}

\author{Adriana de Lorenzo-C\'{a}ceres}
\affiliation{Instituto de Astrof\'{\i}sica de Canarias, C/ V\'{i}a L\'{a}ctea, S/N, E-38205 La Laguna, Tenerife, Spain}

\author{Dmitri A. Gadotti}
\affiliation{Centre for Extragalactic Astronomy, Department of Physics, Durham University, South Road, Durham DH1 3LE, UK}

\author{Jairo M\'{e}ndez-Abreu}
\affiliation{Instituto de Astrof\'{\i}sica de Canarias, C/ V\'{i}a L\'{a}ctea, S/N, E-38205 La Laguna, Tenerife, Spain}
\affiliation{Departamento de Astrof\'{\i}sica, Universidad de La Laguna, Avda. Astrof\'{\i}sico Fco. S\'{a}nchez s/n, 38200, La Laguna, Tenerife, Spain}

\author{John E. Beckman}
\affiliation{Instituto de Astrof\'{\i}sica de Canarias, C/ V\'{i}a L\'{a}ctea, S/N, E-38205 La Laguna, Tenerife, Spain}
\affiliation{Departamento de Astrof\'{\i}sica, Universidad de La Laguna, Avda. Astrof\'{\i}sico Fco. S\'{a}nchez s/n, 38200, La Laguna, Tenerife, Spain}
\affiliation{Consejo Superior de Investigaciones Cient\'ificas, Spain}

\author{Ralf Bender}
\affiliation{Max-Planck-Insitut f\"{u}r extraterrestrische Physik, Giessenbachstrasse, D85748 Garching, Germany}
\affiliation{Universit\"{a}ts-Sternwarte M\"{u}nchen, Scheinerstrasse 1, D-81679 M\"{u}nchen, Germany}

\author{Niv Drory}
\affiliation{McDonald Observatory, The University of Texas at Austin, 1 University Station, Austin, Texas 78712, USA}



\author{Deanne Fisher}
\affiliation{Center for Astrophysics and Supercomputing, Swinburne University of Technology, Hawthorn, Australia}

\author{Ulrich Hopp}
\affiliation{Max-Planck-Insitut f\"{u}r extraterrestrische Physik, Giessenbachstrasse, D85748 Garching, Germany}
\affiliation{Universit\"{a}ts-Sternwarte M\"{u}nchen, Scheinerstrasse 1, D-81679 M\"{u}nchen, Germany}

\author{Matthias Kluge}
\affiliation{Max-Planck-Insitut f\"{u}r extraterrestrische Physik, Giessenbachstrasse, D85748 Garching, Germany}
\affiliation{Universit\"{a}ts-Sternwarte M\"{u}nchen, Scheinerstrasse 1, D-81679 M\"{u}nchen, Germany}

\author{Tutku Kolcu}
\affiliation{Astrophysics Research Institute, Liverpool John Moores University, Twelve Quays House, Egerton Wharf, Birkenhead CH41 1LD, UK}

\author{Witold Maciejewski}
\affiliation{Astrophysics Research Institute, Liverpool John Moores University, Twelve Quays House, Egerton Wharf, Birkenhead CH41 1LD, UK}

\author{Kianusch Mehrgan}
\affiliation{Max-Planck-Insitut f\"{u}r extraterrestrische Physik, Giessenbachstrasse, D85748 Garching, Germany}
\affiliation{Universit\"{a}ts-Sternwarte M\"{u}nchen, Scheinerstrasse 1, D-81679 M\"{u}nchen, Germany}

\author{Taniya Parikh}
\affiliation{Max-Planck-Insitut f\"{u}r extraterrestrische Physik, Giessenbachstrasse, D85748 Garching, Germany}
\affiliation{Universit\"{a}ts-Sternwarte M\"{u}nchen, Scheinerstrasse 1, D-81679 M\"{u}nchen, Germany}

\author{Roberto Saglia}
\affiliation{Max-Planck-Insitut f\"{u}r extraterrestrische Physik, Giessenbachstrasse, D85748 Garching, Germany}
\affiliation{Universit\"{a}ts-Sternwarte M\"{u}nchen, Scheinerstrasse 1, D-81679 M\"{u}nchen, Germany}

\author{Marja Seidel}
\affiliation{IPAC, California Institute of Technology, 1200 East California Boulevard, Pasadena, CA 91125, USA}

\author{Jens Thomas}
\affiliation{Max-Planck-Insitut f\"{u}r extraterrestrische Physik, Giessenbachstrasse, D85748 Garching, Germany}
\affiliation{Universit\"{a}ts-Sternwarte M\"{u}nchen, Scheinerstrasse 1, D-81679 M\"{u}nchen, Germany}

\begin{abstract}
We present photometric and morphological analyses of nuclear star clusters (NSCs) -- very dense, massive star clusters present in the central regions of most galaxies -- in a sample of 33 massive disk galaxies within 20 Mpc, part of the ``Composite Bulges Survey.'' We use data from the \textit{Hubble Space Telescope} including optical (F475W and F814W) and near-IR (F160W) images from the Wide Field Camera 3. We fit the images in 2D to take into account the full complexity of the inner regions of these galaxies (including the contributions of nuclear disks and bars), isolating the nuclear star cluster and bulge components.  We derive NSC radii and magnitudes in all 3 bands, which we then use to estimate NSC masses. Our sample significantly expands the sample of massive late-type galaxies with measured NSC properties.  We clearly identify nuclear star clusters in nearly 80\% of our galaxies, putting a lower limit on the nucleation fraction in these galaxies that is higher than previous estimates.  We find that the NSCs in our massive disk galaxies are consistent with previous NSC mass-NSC radius and Galaxy Mass-NSC Mass relations.  However, we also find a large spread in NSC masses, with a handful of galaxies hosting very low-mass, compact clusters. Our NSCs are aligned in PA with their host galaxy disks but are less flattened.  They show no correlations with bar or bulge properties. Finally, we find the ratio of NSC to BH mass in our massive disk galaxy sample spans a factor of $\sim$300.
\end{abstract}


\section{Introduction} 
\label{sec:intro}

Most galaxies today have a central massive object that is indicative of current or past extreme activity at their centers. These central massive objects can be a supermassive black hole (SMBH), a nuclear star cluster (NSC), or a combination of both.  A review of NSC properties has recently been compiled by \citet{neumayer2020}.  
NSCs are extremely luminous objects that are present in the centers of a majority of all types of galaxies \citep[e.g.,][]{phillips1996,carollo1998,boker2002,cote2006,georgiev2009}. They are compact, massive star clusters with an effective radius ranging from 3 -- 20 pc \citep[e.g.,][]{boker2004, cote2006, walcher2006, georgiev2014, georgiev2016} and dynamical and stellar populaton based masses from $10^{5}$--$10^{8}$~\Msun{} \citep[e.g.,][]{walcher2005, rossa2006, erwin2012, spengler2017, nguyen2018}. NSCs are known to be more massive and denser than GCs \citep[][]{walcher2005, hopkins2010}. 

NSCs are mostly frequently found in intermediate-mass galaxies  ($> 10^{8-10}$ \Msun) with a nucleation fraction (the fraction of galaxies studied that host an NSC) of $\sim 70$--90\% in the early-type galaxies \citep{cote2006, turner2012, denbrok2014,sanchez2019,hoyer2021} and $> 75$\% in the late-types \citep{boker2002,seth2006,georgiev2014,neumayer2020, hoyer2021} in this same mass range. More recent studies of early-type galaxies have shown a strong mass dependence and milder environmental dependence on the nucleation fraction. \citet{sanchez2019}, \citet{zanatta2021}, and \citet{carlsten2022} found the nucleation fraction to be as high as 90\% for galaxy stellar masses of $\sim$10$^9$ M$_\odot$, decreasing towards both lower and higher mass galaxies. \citet{neumayer2020} and \citet{hoyer2021} found a similar trend for late-type galaxies at lower masses, but at higher masses ($> 10^{10}$ \Msun) a lack of data means we do not know if the nucleation fraction of late-types decreases in the same way as early-types.

NSCs are located at the dynamical centers of their galaxies \citep[][]{neumayer2012}. NSCs and SMBHs are known to co-exist, including in the Milky Way \citep[e.g.,][]{seth2008, graham2009, seth2010, georgiev2016, nguyen2019, neumayer2020}. The relative masses of NSCs and SMBHs appears to span a wide range, with NSCs being more massive than SMBHs in many lower mass galaxies, while SMBHs are the dominant component in higher mass galaxies \citep{graham2009, neumayer2020}.
The relationship between NSCs and SMBHs is complicated and unclear.  NSCs may provide a seeding mechanism to create and/or grow massive black holes at the centers of galaxies \citep[e.g.][]{stone2017,inayoshi2020}.  
A possible consequence of NSCs and SMBHs co-existing is the presence of Tidal Disruption Events (TDEs). This occurs when tidal forces pull apart a star as it approaches the SMBH. TDEs are observed mostly in low-mass galaxies with some intermediate-mass galaxies \citep[galaxy stellar masses between 10$^9$--10$^{10}$;][]{wevers2019}. 

The formation history of NSCs can be directly probed through their stellar populations.  Spectroscopic analyses of NSCs show they have have multiple stellar populations, extended star formation histories, and strong rotation \citep[e.g.,][]{walcher2006,rossa2006,seth2006,kacharov2018,pinna2021,fahrion2021, hannah2021, fahrion2022b}.
Another way of studying the formation mechanisms of NSCs is through their correlation with the SMBHs at their centers and their surrounding host galaxies.
Mass measurements of NSCs and host galaxies have shown that these quantities are strongly correlated; initially these were thought to form a relation sismilar to SMBHs \citep[e.g.][]{ferrarese2006,wehner2006,balcells2007,graham2012,scott2013}; 
however, more recent work shows the NSC and SMBH scaling relations are quite different \citep[e.g.][]{erwin2012,georgiev2016,sanchez2019}. 

An in-depth analyses of NSC scaling relations in different galaxy environments can help us understand what physical mechanisms play an important role in NSC formation and how NSCs impact the overall evolution of the host galaxy. Evidence from previous studies \citep[e.g.,][]{hartmann2011, antonini2012, neumayer2020,fahrion2021,fahrion2022} show that there are two primary formation mechanisms that drive the growth of NSCs, (1) star cluster mergers \citep[][]{tremaine1975, gnedin2014}, or (2) \textit{in situ} formation \citep[][]{bekki2007, antonini2015}. The relative importance of these mechanisms seems to depend on the galaxy stellar mass. Specifically, recent work \citep{neumayer2020, fahrion2021,fahrion2022,fahrion2022b} finds that the NSCs of low mass galaxies primarily grow from cluster mergers, while in higher mass galaxies the NSCs grow from \textit{in situ} formation. A reflection of this transition seems to be present in the scaling relations as well, with the NSC masses scaling with the square root of their host galaxy masses at lower masses but steepening to a more linear relationship in higher mass galaxies \citep{denbrok2014,sanchez2019}.

The changing properties of NSCs as a function of galaxy morphology and the resulting implications for their formation are not yet fully understood. Overall, there have been more studies of early-type than late-type galaxies, especially towards higher masses where dust and bulge contributions make studying late-type NSCs more challenging.  Therefore, most observations of NSCs in late-type galaxies have focused on lower-mass galaxies; this includes both HST imaging to quantify structure \citep{georgiev2009, boker2002,boker2004,georgiev2014,carson2015,hoyer2023}, and spectroscopic observations focused on kinematics and stellar populations \citep{walcher2005,kacharov2018,pinna2021,fahrion2022b}. These observations are broadly consistent with the NSC mass trends discussed in the previous paragraph. However, some differences have been suggested in NSCs in early vs.~late-type galaxies.  For instance, the compilation of data by \citet{georgiev2016} found that the sizes of NSCs in massive early-type galaxy are $\sim$2$\times$ larger than those in late-type galaxies, although \citet{neumayer2020} suggested this difference may only exist at the highest masses.  Late-type galaxies also seem to show stronger rotation on average than early-type galaxies \citep{pinna2021}.

Some studies exist of higher-mass, more bulge-dominated late-type galaxies as well, most notably the studies \citet{carollo1998,carollo2002}, which found that a a majority of massive spiral galaxies do host NSCs.  However, the information available on these NSCs is quite heterogeneous (i.e.~photometric bands, sizes) making them challenging to interpret together. Stellar population measurements of a subset of these galaxies by \citet{rossa2006} found that these galaxies clusters tended to be older than those in lower mass late-type galaxies.  The recent paper by \citet{hoyer2023b} shows the promise of JWST for studying NSCs in massive galaxies due to its high angular resolution, ability to penetrate dust, and the broad spectral energy distribution measurements it can obtain. Despite these studies, we still know little about the populations of NSCs in massive late-type galaxies (like the Milky Way; MW hereafter), which leads to a lack of knowledge about the NSCs in these galaxies and their co-existence with the ubiquitous SMBHs in these galaxies. 
In this paper, we focus on NSCs in high-mass ($\log (\Mstar/\Msun) > 10$) late-type galaxies, presenting a study of NSCs in 33 galaxies with high resolution, uniform imaging from the Composite Bulges Survey (CBS; Erwin et al. in prep).

Section~\ref{sec:data} describes the galaxy sample selected from CBS, the high resolution HST data, and the other data used in this work. In Section~\ref{sec:modelling}, we describe in detail the morphological modelling process for the galaxies and NSCs, including derivation of  morphological NSC parameters. We also discuss the quality of the fits and error estimates. In Section~\ref{sec:nscprops}, we focus on the properties of the NSCs, including the nucleation fraction of our sample, NSC magnitudes and masses, and correlations between NSC internal properties. In section \ref{sec:nscgalrelations}, we present scaling relations of the NSCs with their host galaxies. We also discuss briefly the presence of SMBHs in our galaxy sample. Section \ref{sec:conclusion} summarizes our work in this paper.

\section{Data sample}
\label{sec:data}
The work shown in this paper is part of the Composite Bulges Survey (CBS; Erwin et al. in prep). This survey is aimed at a detailed analysis of the stellar morphology and populations in the inner 1--2 kpc of massive disk galaxies, using both \textit{HST} optical and near-IR imaging and VLT-MUSE IFU data.
\begin{figure}[!t]
\advance\leftskip-1cm
\includegraphics[trim=0cm 1cm 0cm 2.5cm, clip, scale=0.36]{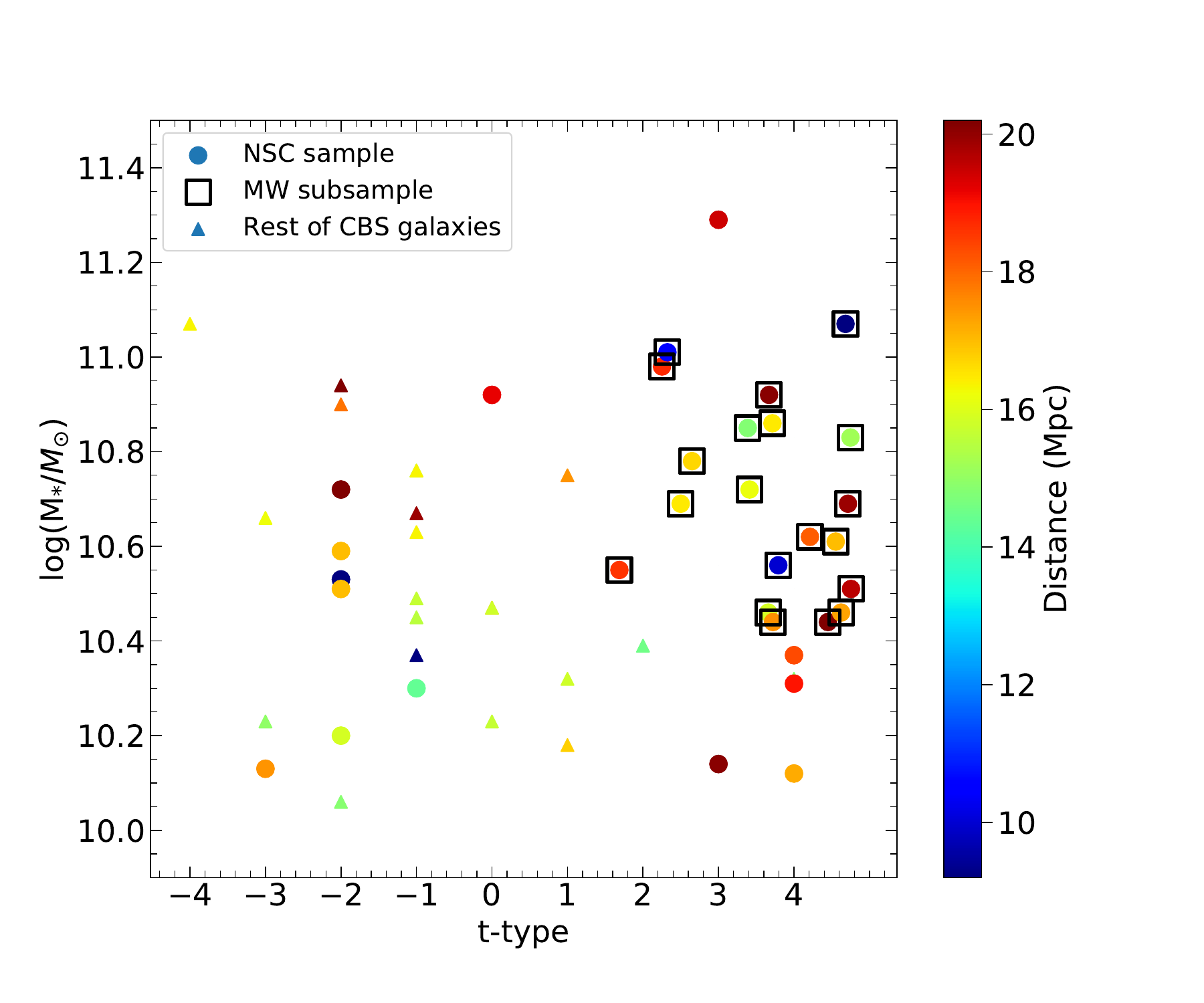}
\caption{The stellar mass, distance, and Hubble type of the 33 NSC sample galaxies (circles) selected from the 53 galaxy CBS survey (triangles). The MW-like subsample of 20 galaxies are shown as squares.}
\label{fig:cbsdata}
\end{figure}

CBS is based on a mass- and volume-limited sample of disk galaxies.  These galaxies were selected from the \citet[][RC3]{RC3} catalog, restricted to galaxies with distances $\leq 20$ Mpc, stellar masses $\geq 10^{10}$ \Msun, S0--Sbc morphologies, inclinations between 35$^{\circ}$ and 60$^{\circ}$, declinations $\delta \leq +20^{\circ}$, and galactic latitudes $|b| > 20^{\circ}$. Based on these selections, a total of 53 galaxies were selected; these include a large number of Virgo Cluster galaxies.  

\subsection{NSC and MW-like sample selection}
\label{subsec:MW-likedata}
For the work shown in this paper we present NSC fits for a sub-sample of 33 galaxies from the parent CBS sample (hereafter the ``NSC sample''). This sub-sample of the CBS survey was chosen to include a complete set of 20 galaxies form a complete Milky-Way like sample (hereafter the ``MW-like'' subsample). Our MW is a special galaxy which hosts the nearest NSC that can be studied in unparalleled detail. Understanding how typical or atypical the NSC in the MW is with the NSCs in similar galaxies is important. Hence we create this sample of MW-like galaxies which represents all galaxies having spiral morphologies ($t \geq 1$) and stellar masses from $10^{10.4}$ to $10^{11.1}$ \Msun{} from CBS. We present NSC fits for the complete sample of CBS galaxies that have fit these criteria. In addition to these galaxies, we fit the NSCs of 13 additional galaxies  -- these galaxies are a random subsample of galaxies for which we had available larger scale models; the full sample of CBS galaxies morphological fits will be presented in Erwin et al.\ in prep.
The circles in Figure \ref{fig:cbsdata} show all the 33 of the NSC sample galaxies; of these the MW-like subsample are shown as squares and the rest of the CBS sample as triangles. {\em As can be seen, a majority of the CBS galaxies that are not included in our NSC sample} are S0 type galaxies, while the NSC sample represents a nearly complete set of the Sa-Sbc galaxies. The properties of the NSC sample galaxies are summarized in Table \ref{tab:galdata}.

\subsection{HST data from CBS}
\label{subsec:hstdata}
For modeling the central morphology of each galaxy, we use a consistent set of high resolution HST data obtained using the Wide Field Camera 3 (WFC3) in the UVIS and IR modes (Cycle 25, Proposal ID 15133). We use the full-field WFC3/IR image in the F160W filter with a total integration time of 600s. The optical images were restricted to the C1K1C aperture for efficient readout and data transmission. This aperture is a 1024 X 1024-pixel subarray of the full-field WFC3/UVIS images centered on the galaxy nuclei. We use the F814W filter with a total integration time of 500s and the F475W filter with a total integration time of 700s. All of our images are divided into four dithered exposures to provide sub-pixel sampling. We note that the galaxies more than fill both detectors in many cases.  

All the individual exposures from each band are combined using the Python-based DrizzlePac code. We set the output image scale for the UVIS images to 0.03\arcsec/pixel and for the IR images to 0.06\arcsec/pixel with a pixfrac = 0.7 in all cases. The sky subtraction during the processing of these images is turned off; we estimate the sky background using larger-scale ground-based or \textit{Spitzer} IRAC1 images as explained in detail in Section \ref{subsec:skysubtraction}. 

\subsection{Other data used in this work}
\label{subsec:otherdata}
For the large scale galaxy fitting and to determine the F160W image sky levels, we use \textit{Spitzer} IRAC1 (3.6\micron) images. The images for the galaxies come from either the S$^4$G survey with a final mosaic pixel size of 0.75\arcsec/pixel \citep[][]{sheth2010} or  archive-generated mosaic images \citep[][]{watkins2022} for
galaxies not in S$^4$G, with a default archive mosaic pixel of 0.6\arcsec/pixel. The program IDs for the IRAC1 images are listed in Table \ref{tab:galdata}.

For determining the background sky levels in the UVIS images (Section~\ref{subsec:skysubtraction}), we use either SDSS $g$ and $i$ images or (for galaxies without SDSS images) $B$, $V$, and $I$ images from the Carnegie-Irvine Galaxy Survey \citep{ho2011}, kindly provided by Luis Ho.  

\begin{deluxetable*}{cccccccccccc}
\tabletypesize{\scriptsize}
\tablewidth{-10pt}
\tablecolumns{8}
\tablecaption{\label{tab:galdata}Properties of galaxies in the NSC sample}
\tablehead{\colhead{Galaxy} & \colhead{Hubble type} & \colhead{RC3 type} & \colhead{Sample} & \colhead{log(~M$_\star$/~M$_\odot$)} & \colhead{Distance} & \colhead{Source} & \colhead{Spitzer program ID} & A$_{F814W}$ & \colhead{PA$_{\rm Gal}$} & \colhead{$\epsilon_{\rm Gal}$} & \colhead{log(M$_{BH})$}\\
\nocolhead{} & \nocolhead{} & \nocolhead{} & \nocolhead{} & \nocolhead{} & \colhead{(Mpc)} & \nocolhead{} & \nocolhead{} & \nocolhead{} & \colhead{degrees} & \nocolhead{} & \colhead{M$_{\odot}$}\\
\colhead{(1)} & \colhead{(2)} & \colhead{(3)} & \colhead{(4)} & \colhead{(5)} & \colhead{(6)} & \colhead{(7)} & \colhead{(8)} & \colhead{(9)} & \colhead{(10)} & \colhead{(11)} & \colhead{(12)}}
\startdata
IC 2051 & 4 & SBbc & MW-like & 10.69 & 19.9 & 1 & 61060 & 0.18 & 70 & 0.42 & --\\
NGC 289 & 4 & SBbc & MW-like & 10.44 & 20.2 & 1 & 61064 & 0.03 & 130 & 0.29 & --\\
NGC 613 & 4 & SBbc & MW-like & 10.62 & 18.1 & 1 & 61064 & 0.03 & 115 & 0.19 & --\\
NGC 1097 & 3 & SBb& MW-like & 10.85 & 14.8 & 1 & 159 & 0.04 & 134 & 0.29 & --\\
NGC 1300 & 4 & SBbc & MW-like & 10.51 & 19.6 & 1 & 61065 & 0.05 & 87 & 0.17 & 7.91\\
NGC 1440 & -2 & SB0 & NSC & 10.72 & 19.9 & 1 & 10043 & 0.16 & 25 & 0.20 & --\\
NGC 1566 & 4 & SABbc & MW-like & 10.61 & 17.0 & 1 & 159 & 0.01 & 34 & 0.14 & --\\
NGC 2775 & 2 & SAab & MW-like & 10.98 & 18.7 & 1 & 69 & 0.07 & 165 & 0.21 & --\\
NGC 3351 & 3 & SBb & MW-like & 10.56 & 10.0 & 3 & 159 & 0.04 & 10 & 0.29 & --\\
NGC 3368 & 2 & SABab & MW-like & 11.01 & 10.5 & 3 & 69 & 0.04 & 172 & 0.37 & 6.88\\
NGC 3412 & -2 & SB0 & NSC & 10.53 & 11.0 & 2 & 10043 & 0.04 & 152 & 0.45 & 6.85\\
NGC 4237 & 4 & SAbc & NSC & 10.31 & 18.9 & 5 & 50128 & 0.05 & 106 & 0.38 & --\\
NGC 4321 & 4 & SABbc & MW-like & 10.83 & 15.2 & 4 & 159 & 0.04 & 152 & 0.14 & --\\
NGC 4377 & -3 & SAB0 & NSC & 10.13 & 17.7 & 5 & 10043 & 0.06 & 5 & 0.20 & --\\
NGC 4380 & 3 & SAb & MW-like & 10.46 & 15.9 & 5 & 30496 & 0.04 & 157 & 0.46 & --\\
NGC 4450 & 2 & SABab & MW-like & 10.78 & 16.7 & 6 & 159 & 0.04 & 170 & 0.31 & --\\
NGC 4501 & 3 & SAb & MW-like & 10.86 & 16.5 & 6 & 30945 & 0.06 & 141 & 0.49 & 7.30\\
NGC 4531 & -1 & SA0 & NSC & 10.30 & 15.2 & 5 & 61060 & 0.07 & 154 & 0.33 & --\\
NGC 4548 & 3 & SBb & MW-like & 10.72 & 16.2 & 3 & 3674 & 0.06 & 149 & 0.23 & --\\
NGC 4578 & -2 & SA0 & NSC & 10.20 & 16.4 & 5 & 10043 & 0.03 & 31 & 0.30 & --\\
NGC 4579 & 3 & SABb & MW-like & 10.92 & 20.1 & 5 & 159 & 0.06 & 95 & 0.22 & --\\
NGC 4608 & -2 & SB0 & NSC & 10.69 & 17.3 & 5 & 10043 & 0.03 & 105 & 0.18 & --\\
NGC 4612 & -2 & SAB0 & NSC & 10.51 & 17.3 & 5 & 10043 & 0.04 & 143 & 0.27 & --\\
NGC 4643 & 0 & SB0/a & NSC & 10.92 & 19.1 & 1 & 61063 & 0.05 & 53 & 0.20 & --\\
NGC 4689 & 4 & SAbc & NSC & 10.12 & 17.5 & 5 & 69 & 0.04 & 163 & 0.20 & --\\
NGC 4698 & 2 & SAab & MW-like & 10.69 & 16.5 & 7 & 30496 & 0.04 & 170 & 0.50 & --\\
NGC 4699 & 3 & SABb & NSC & 11.29 & 19.3 & 1 & 61064 & 0.05 & 35 & 0.19 & 8.24\\
NGC 5121 & 1 & SAa & MW-like & 10.55 & 18.6 & 1 & N/A & 0.11 & 28 & 0.22 & --\\
NGC 5248 & 4 & SABbc & MW-like & 10.46 & 17.3 & 1 & 69 & 0.04 & 114 & 0.28 & --\\
NGC 5364 & 4 & SAbc & NSC & 10.37 & 18.4 & 1 & 61065 & 0.04 & 37 & 0.28 & --\\
NGC 6744 & 4 & SABbc & MW-like & 11.07 & 9.2 & 4 & 10136 & 0.07 & 14 & 0.37 & --\\
NGC 7177 &  3 & SABb & MW-like & 10.44 & 17.5 & 1 & 30496 & 0.11 & 83 & 0.31 & --\\
NGC 7513 & 3 & SBb & NSC & 10.14 & 19.8 & 1 & 61065 & 0.06 & 105 & 0.29 & --\\
\enddata
\tablecomments{(1) Galaxy name, (2) morphological Hubble classification type, (3) Galaxy classification from the RC3 catalog \citep{RC3}, (4) NSC sample or MW-like sample defined in this work, (5) logarithmic galaxy stellar mass derived using the HyperLEDA $B$ band absolute magnitude, ($B-V$) color and  M/L ratio from \citet[][]{bell2003}, (6) Distance in Mpc, (7) Source for distances: 1 = Virgocentric-corrected redshift from HyperLEDA + H0 = 72; 2 = Surface Brightness Fluctuation (SBF; Jensen+2021); 3 = Surface Brightness Fluctuation (SBF; Tonry+2001, with correction from Mei+2005); 4 = Cepheids (metallicity-corrected values, Freedman+2001); 5 = TRGB (Anand+2021); 6 = SBF (Cantiello+2018); 7 = Default Virgo Cluster distance, (8) Spitzer ID program, (9) F814W galactic extinction obtained from Ned IPAC. For the F475W image and F160W images, we scale the extinction by 0.49 and 2.87 respectively, (10) Galaxy Position Angle (PA), (11) Galaxy Ellipticity, (12) Black Hole mass from \citet{saglia2016} used in Section~\ref{subsec:nscbhmass}.}
\end{deluxetable*}

\section{Dissecting the Galaxies' Stellar Components}
\label{sec:modelling}
In this section we explain in detail the process of modeling each of the galaxy components for the NSC sample galaxies. We describe how we identify and create accurate models for the inner regions of the galaxy, especially the NSC component. 
The fits for each of the galaxies is done as a three-step process. First, we use the sky-subtracted IRAC1 images to fit the large-scale components (mainly the bar and disc related components) as explained in section \ref{subsec:iracfits}. Using this fit, we then use the HST images (first the wider field-of-view (FOV) F160W images and then the UVIS images) to fit for the inner components (e.g., bulge, nuclear disc). Once we have a good model for the overall structure of the galaxy, we then identify, refine and constrain the NSC component using mostly the HST UVIS images as explained in section \ref{subsec:hstfits}. 

We fit our galaxies using the fast, multi-component image fitting program \textsc{imfit}\footnote{\url{https://github.com/perwin/imfit}} version 1.8 \citep{imfit}. This code creates 2-D models for each galaxy component using user-defined input parameters, adds them together, and then convolves the summed image with a user-supplied PSF image. These are fitted to the images with the default $\chi^2$ minimization using a Levenberg-Marquardt (L-M) minimization algorithm.  In some cases where we run into local minima issues while modeling the images, we also use the Nelder-Mead (N-M) minimization algorithm to explore a wider range of solutions. The per-pixel uncertainties are estimated from the data values using the Gaussian approximation to Poisson statistics. When fitting models to the IRAC1 images, we convert the pixel values in the latter to ADUs with an assumed A/D gain of 3.7; the units of the HST images are in electrons and require no scaling. 

\begin{figure*}[!t]
\includegraphics[trim=0cm 25cm 0cm 0cm, clip, scale=0.21]{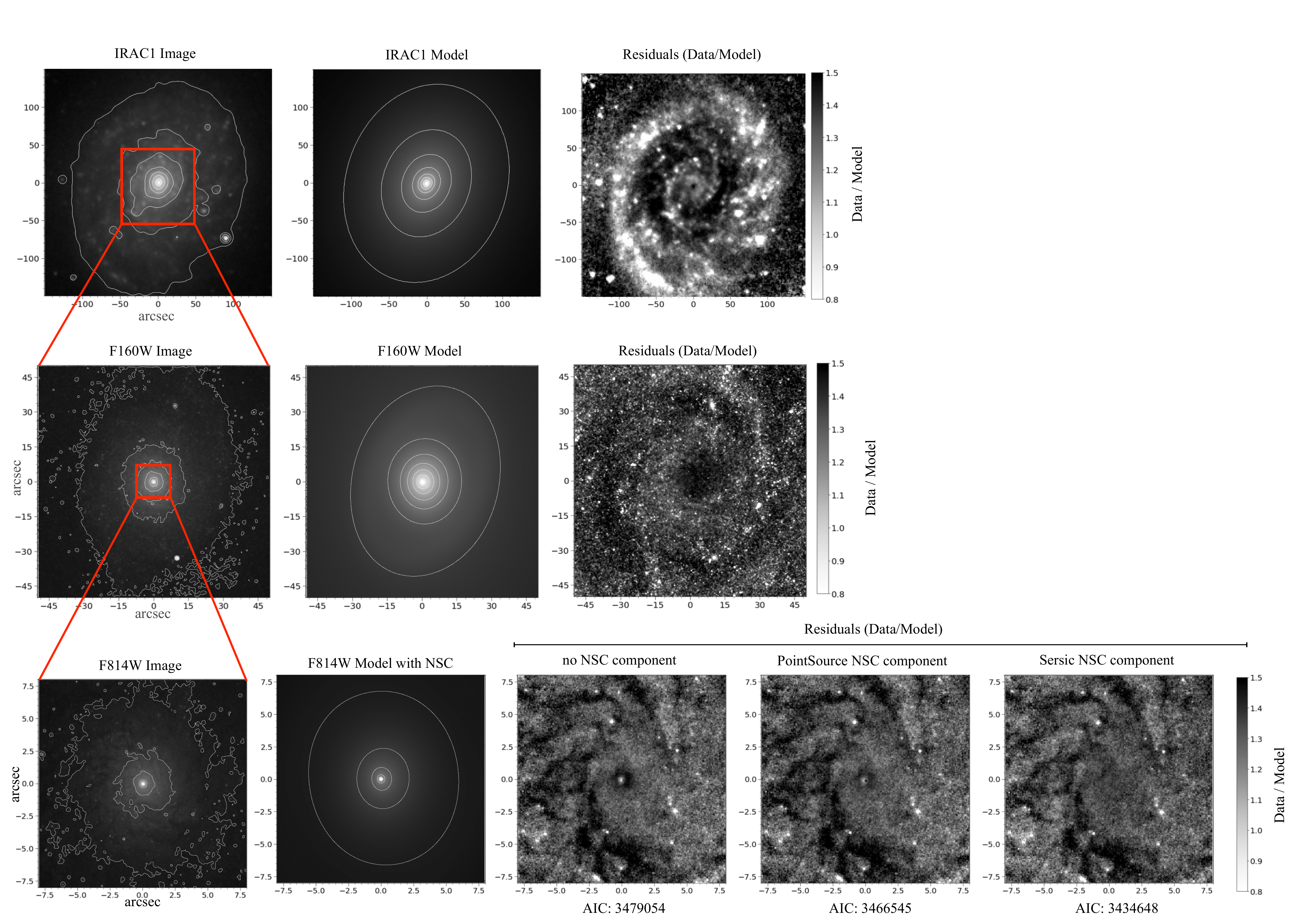}
\caption{2D image modeling of an example galaxy, NGC 4689, using \textsc{imfit}. Each row describes fits on a different scale; the top row shows the largest scale fits to the \textit{Spitzer} IRAC1 image, the middle row, the HST/WFC3-IR F160W image, while the bottom row shows the smallest scale and highest resolution fits to the HST/WFC3-UVIS F814W image.  The left column shows the data, the next column rigth shows the best-fit model, and the right most columns show the residuals.  In the bottom row, three residual images are shown, the left one is from a fit with no NSC, the middle with a point source NSC component, and the right with the best-fit model image shown; the stretch of all three of these residual images is identical and clearly shows the need for a resolved NSC component in this galaxy.}
\label{fig:modelingprcoess}
\end{figure*}

\begin{figure*}[ht!]
\includegraphics[trim=0.75cm 0.5cm 0.5cm 0.5cm, clip, scale=0.48]{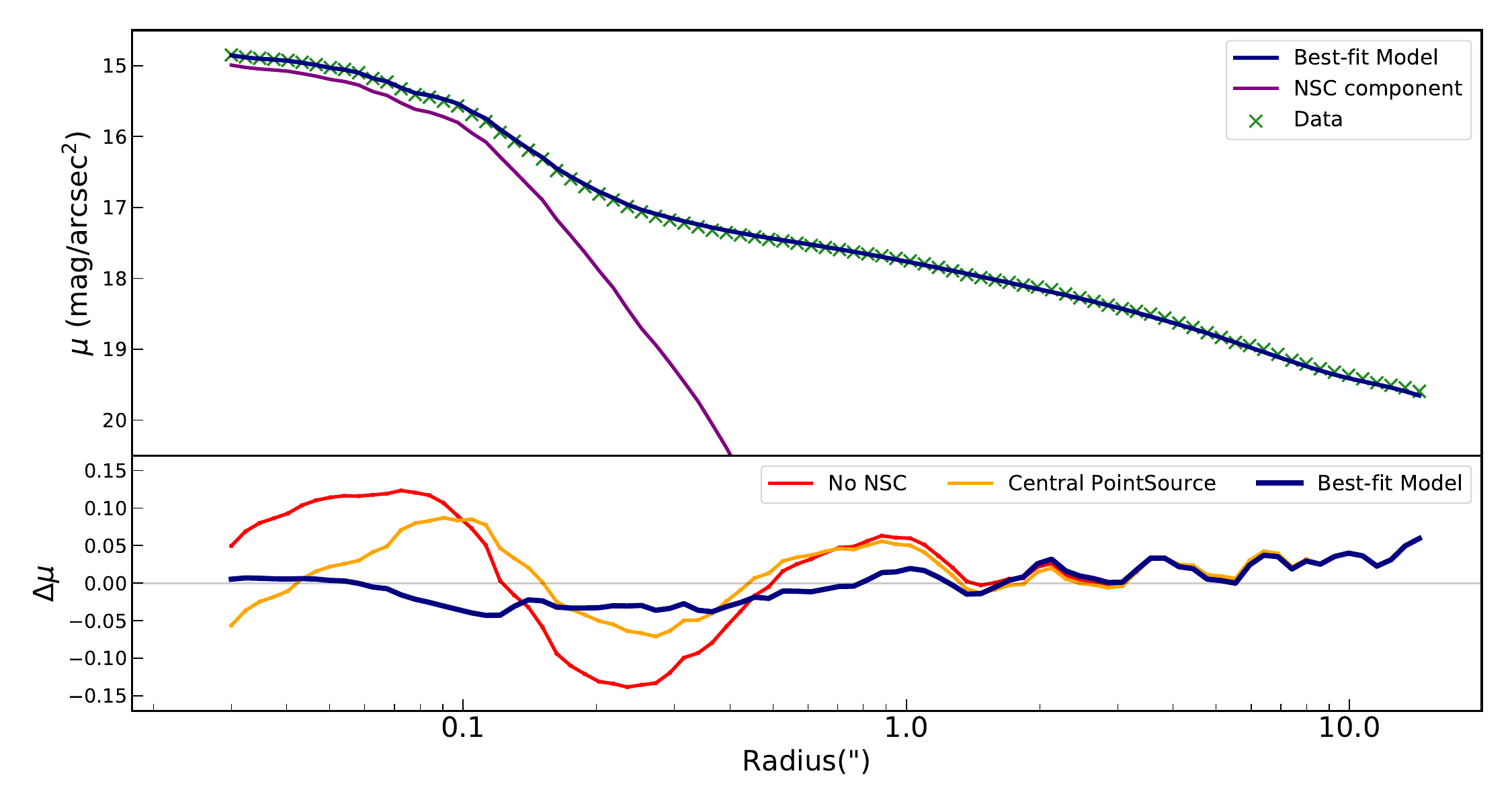}
\caption{\textbf{Top panel:} An example 1-D surface brightness profile of NGC 4689 from our F814W imaging and modeling. Note this is the same galaxy shown in Fig.~\ref{fig:modelingprcoess}.  The data is shown as green crosses, the best-fit model is shown as the blue solid line, and the best-fit NSC component is shown as the purple solid line.  \textbf{Lower panel:} The residuals (model -- data) in magnitudes.  Three different residuals are shown corresponding to the right three panels in Fig.~\ref{fig:modelingprcoess}.  The best-fit model with a resolved NSC, point source NSC model, and no NSC model are shown as blue, orange, and red lines respectively.} 
\label{fig:sbprofile}
\end{figure*}

We use a wide range of 2-D functions to accurately fit our galaxies disk, bulge and bar components as well as their NSCs.    The most commonly used components are listed in detail in Appendix~\ref{app:components}.

\subsection{Sky subtraction}
\label{subsec:skysubtraction}
While the primary goal of this paper is the nuclear morphology of galaxies, our modeling still requires accurate sky background estimates so that we can correctly model the large scale components of the glaaxies.  The galaxies in our sample are massive and nearby,  and thus they are almost always larger in angular size than the FOV of our HST images. (This is nearly always true for the $160 \times 160\arcsec$ FOV of WFC3-IR, and always true for our $41 \times 41\arcsec$ WFC3-UVIS C1K1C images.) It is thus not possible to accurately estimate the sky background from the HST images themselves. To determine reasonable estimates for the sky backgrounds, we match surface-brightness profiles from ellipse-fitting of the HST images to those from ground- or spaced-based images at similar wavelengths with larger FOV (i.e., large enough to determine the overall background outside the galaxy). The surface brightnesses were measured with the \textsc{iraf} \texttt{ellipse} task using ellipses with position angles and ellipticities matching the galaxy main-disk orientation; on the non-HST images, we used masks reproducing the orientation and FOV of the HST images. The resulting HST profile was then scaled to match the corresponding larger-FOV profile, using measurements outside the region strongly affected by differences in the PSFs, including an additive component representing the unknown HST background.

For the UVIS F475W and F814W images, the reference images are SDSS $g$ and $i$-band images, respectively, where we apply a mask mimicking the orientation and FOV of the HST images. In the case of galaxies lacking SDSS images, we make use of images from the Carnegie-Irvine Galaxy Survey \citep{ho2011}, matching their $I$-band images to the F814W images and averaging the results from matching the $B$- and $V$-band images to the F475W images. For a small number of galaxies where the matching with ground-based profiles failed  we use the average sky values obtained from the rest of the galaxies in the NSC sample: 9.14 electrons for F814W images and 13.56 electrons for the F475W images. For the HST F160W images we generally used Spitzer IRAC1 images as the reference images. (Exceptions were NGC~3351, where we used an $H$-band image from the 2MASS Large Galaxy Atlas \citep{jarrett2003}, and NGC~1300 and NGC~4321, where we made use of $H$-band images from \citet{grosbol2012}.) Just as with the optical images, we match the surface brightness profiles within the larger HST F160W FOV to that of the IRAC1 images. (See Erwin et al., in prep, for more details.)

The sky background is incorporated into our \textsc{imfit} modeling using a constant (fixed value) FlatSky function. 

\subsection{Large scale fitting}
\label{subsec:iracfits}
We initially use the large scale IRAC1 images to fit the larger, outer components of the galaxy. For all of our galaxies, we started with the combination of a single S{\'e}rsic component and a single Exponential component, and then added components to best fit the galaxy.  
These models were fit using an appropriate PSF; we have used the official in-flight Pixel-Response-Function (PRF) images\footnote{\url{https://irsa.ipac.caltech.edu/data/SPITZER/docs/irac/ calibrationfiles/psfprf/}} that are down-sampled to the appropriate pixel scale from the original scale of 0.24\arcsec/pixel at~column,~row~=~129,129; the approximate central location of most of our galaxies. We also generate a bad-pixel mask to remove bright foreground stars and background galaxies; we do this by running SExtractor \citep{SExtractor1996} on the image, and then scaling detected objects using circles (for stars) or ellipses (for background galaxies). Notable image defects are masked by hand.

After running our initial fit, we then refine the fit by adding components based on (1) visible patterns in the residual images, (2) surface brightness profiles, and (3) changes in the ellipticity and position angle of the galaxy isophotes. For some galaxies, we find significantly better fits if we replace the initial Exponential with a BrokenExponential component. (Figures 5 and 8 in \citet{erwin2021} show examples of this for the galaxies NGC~4608 and NGC~4643.) We also incorporate information from the galaxy morphology (e.g.~known rings or bars) from previous studies.  We continue to add additional components to the model until the galaxy is well represented; specifically, when there are no clearly visible systematic residual patterns that could be fit with an additional component.  Typically, this includes an exponential or broken exponential disk component, a bar component, and at least one bulge component.  Occasionally, we also fit ring and spiral arm components, especially when this is critical for understanding the bar and bulge structures.  These components are always an addition to the initial S{\'e}rsic + Exponential model, except for cases where we have replaced the initial Exponential with a BrokenExponential.

An example of this fitting procedure can be seen in the upper panel of Figure~\ref{fig:modelingprcoess} which shows the IRAC1 image, best-fitting model and the residual ratio image for NGC 4689.

\subsection{HST image fitting}
\label{subsec:hstfits}
Once we have the best-fit IRAC1 model, we then model the galaxy's inner components in more detail using the HST images, beginning with the F160W image.  This includes components such as a boxy-peanut bulge, classical bulge, star forming disk, etc..\\

\noindent {\em PSFs --} For each model we fit to the HST data (in all three band images), we provide an appropriate PSF image for convolution. For the HST images, we generated the PSF images using the \texttt{grizli} software\footnote{\url{https://github.com/gbrammer/grizli}}. An ``empirical PSF'' image from \citet{anderson2016}\footnote{\url{http://www.stsci.edu/~jayander/STDPSFs/WFC3IR/}} is inserted into the each of the four individual exposures at the location of the galaxy center; these are then run through the same drizzling process used to prepare the final HST image (as explained in section \ref{subsec:hstdata}). The final PSF image is then extracted from the combined, drizzled image.  \\

\noindent {\em Masks --} Foreground objects and dust can prevent us from getting an accurate model fit to the data.  We initially identify the foreground stars, background galaxies and other image defects using SExtractor \citep{SExtractor1996}. We then create a mask to flag all those pixels in the image. We also mask other regions of the galaxies that are not well fit by our models, these include spiral arms in many cases, as well as other non-axisymmetric features that are challenging to model.

Dust extinction near the centers of our galaxies can also significantly impact the best-fit nuclear models.  To mask dust features, we use UVIS F475W-F814W color maps to find reddened pixels. We mask pixels using the distribution of pixel values in unreddened regions, choosing a slightly different color threshold for each galaxy.  We then translate this mask to the F160W image as well.  In addition to masking reddened regions, we also mask star-forming regions in some cases using a blue color threshold.  \\

\noindent {\em Fitting the F160W images --} We start the modeling of the F160W image by using the best-fitting IRAC1 model as an initial guess (this involves translating size and angular parameters to account for differences in the pixel scales and image orientation, and estimating the difference in intensity values for intensity parameters).

Since some galaxy components (i.e. the disk) from the IRAC1 model can extend well outside the F160W FOV, we model these components by holding the translated best-fit shape, orientation and size parameters fixed, and fit only for the intensity of these largest components. For components that are mostly or entirely within the F160W [or WFC3-IR] FOV, we use the IRAC1 values as initial guesses, but leave all parameters free. We then iterate by running \textsc{imfit} to determine the best-fit parameters, followed by inspection of the residuals, adding additional components to the model if needed and re-running the fit.
 
In addition to examining the residuals, we also apply a more quantitative approach to adding components.  Specifically, we add components when (1) we see residuals in the radial profile that are $> 10$\% of the data value and when (2) the addition of a component improves (i.e., reduces) the Akaike Information Criterion \citep[$\Delta$AIC;][]{akaike1974} of the fit by 1000 or more relative to the original fit \citep[as in][]{erwin2021}.  Often at this stage we include an initial NSC component that fits for the central excess light; we describe our final NSC determinations below.  

An example of this fitting procedure can be seen in the middle panel of figure \ref{fig:modelingprcoess} which shows the F160W image, best-fitting F160W model image and the residual ratio image for NGC~4689. We obtain a decent F160W model for this galaxy with a preliminary NSC component that fits for the central excess light.  \\

\noindent {\em Translation to the F814W images --}  The F814W images are read out in a sub-array, and thus they span a smaller region (the inner $41\arcsec \times 41\arcsec$) of the galaxy than the F160W images.  They are also higher resolution than the F160W images, with an original pixel scale of 0.04$\arcsec$ and a final processed image scale of 0.03$\arcsec$ pixels (critically sampling the $\sim$0$\farcs$07 FWHM PSF) 
and thus provide better constraints on the nuclear structure than the F160W images as long as the galaxies are not too dusty.  We translate the best-fit F160W model to create the initial guess for the F814W model by (1) scaling the intensity parameters by the ratio of fluxes in a fixed angular area in the two images, and (2) scaling the sizes of components to the higher resolution pixel scale. For components that extend well beyond the edge of our chip, we hold their shape parameters fixed and fit just for their intensity parameters.  
As with the F160W images, we add components as necessary in the higher resolution F814W images; again this often includes an initial NSC fit to the central light excess.  \\

\noindent {\em  Final NSC fitting --} Once we have a good fit to the model that accurately represents the whole F814W band image, we focus on modeling the NSC component.  We choose to use a S{\'e}rsic function to describe the NSC component, following previous work by several authors \citep[e.g.][]{graham2009,carson2015,nguyen2017}.

As a first step, we check for the presence of a nuclear excess in the galaxy. For this, we exclude any NSC component from the F814W model and use the remaining components as initial conditions for a new fit -- this allows us to check whether any nuclear light can be fit through adjusting the parameters of the larger components. We inspect the residuals of these fits and the surface brightness profiles in the central 10-15 pixels for the presence of the excess light.  In all galaxies we find there is central excess light; in many cases this is due to an NSC.  However, we expect central light excess can also be due to an AGN component, in which case we expect the light to be unresolved.  Therefore, as a next step, we add a PointSource component to the model in an attempt to fit the nuclear excess. 

\begin{deluxetable*}{ccccccl}[t!]
\centering
\label{tab:nscdata}
\tablecaption{NSC modeling properties.}
\tablecolumns{7}
\tablehead{\colhead{Galaxy} & \colhead{Sample} & \colhead{Primary Band} & \colhead{${\#}$} & \colhead{$\Delta$~AIC} & \colhead{Quality} & \colhead{Notes}\\
\colhead{(1)} & \colhead{(2)} & \colhead{(3)} & \colhead{(4)} & \colhead{(5)} & \colhead{(6)} & \colhead{(7)}}
\startdata
IC 2051 & MW-like & F160W & 6 & 3.04x10$^{5}$ & 2 & Dust in UVIS affecting NSC fits\\
NGC 289 & MW-like & -- & -- & -- & 0 & Very dusty, bad fit NSC\\
NGC 613 & MW-like & F160W & 6 & 1.01x10$^{2}$ & 1 & Unresolved NSC, weak AGN emissions\\
NGC 1097 & MW-like & F160W & 7 & 3.52x10$^{4}$ & 2 & Dust in UVIS affecting NSC fits\\
NGC 1300 & MW-like & F814W & 6 & 5.03x10$^{2}$ & 1 & Unresolved NSC, weak AGN emissions\\
NGC 1440 & NSC & F814W & 6 & 2.41x10$^{2}$ & 1 & Unresolved NSC\\
NGC 1566 & MW-like & -- & -- & -- & 0 & Center saturated with strong AGN emission\\
NGC 2775 & MW-like & F814W & 5 & 1.11x10$^{3}$ & 1 & Unresolved NSC\\
NGC 3351 & MW-like & F814W & 7 & 3.48x10$^{4}$ & 4 & \\
NGC 3368 & MW-like & F814W & 8 & 3.61x10$^{5}$ & 4 & \\
NGC 3412 & NSC & F814W & 7 & 4.18x10$^{4}$ & 5 & \\
NGC 4237 & NSC & F814W & 3 & 4.65x10$^{5}$ & 4 & \\
NGC 4321 & MW-like & F160W & 10 & 1.22x10$^{5}$ & 2 & Dust in UVIS affecting NSC fits\\
NGC 4377 & NSC & F814W & 6 & 1.24x10$^{4}$ & 5 & \\
NGC 4380 & MW-like & F160W & 6 & 1.75x10$^{4}$ & 2 & \\
NGC 4450 & MW-like & F814W & 9 & 1.08x10$^{4}$ & 4 &\\
NGC 4501 & MW-like & F160W & 3 & 4.71x10$^{5}$ & 2 & Dust in UVIS affecting NSC fits\\
NGC 4531 & NSC & F814W & 5 & 1.42x10$^{5}$ & 4 & \\
NGC 4548 & MW-like & F160W & 7 & 1.64x10$^{4}$ & 2 & Dust in UVIS affecting NSC fits\\
NGC 4578 & NSC & F814W & 5 & 5.19x10$^{4}$ & 5 & \\
NGC 4579 & MW-like & F814W & 5 & 4.92x10$^{4}$ & 2 & Dust in UVIS affecting NSC fits\\
NGC 4608 & NSC & F814W & 6 & 4.12x10$^{5}$ & 5 & \\
NGC 4612 & NSC & F814W & 7 & 5.81x10$^{3}$ & 3 & \\
NGC 4643 & NSC & F814W & 6 & 4.74x10$^{5}$ & 4 & \\
NGC 4689 & NSC & F814W & 5 & 4.32x10$^{3}$ & 4 & \\
NGC 4698 & MW-like & F814W & 7 & 7.28x10$^{3}$ & 5 & \\
NGC 4699 & NSC & F814W & 9 & 1.24x10$^{5}$ & 4 & \\
NGC 5121 & MW-like & F814W & 7 & 2.17x10$^{3}$ & 4 & \\
NGC 5248 & MW-like & F814W & 7 & 2.29x10$^{4}$ & 5 & \\
NGC 5364 & NSC & F814W & 6 & 3.21x10$^{3}$ & 4 & \\
NGC 6744 & MW-like & F160W & 5 & 4.05x10$^{4}$ & 2 & Dust in UVIS affecting NSC fits\\
NGC 7177 & MW-like & -- & -- & -- & 0 & Very dusty, bad fit NSC\\
NGC 7513 & NSC & F814W & 5 & 4.76x10$^{5}$ & 4 & \\
\enddata
\tablecomments{(1) Galaxy name, (2) NSC sample or MW-like subsample, (3) primary band used for modelling the different components in the galaxy, (4) total number of components we fit for in the galaxy (including the NSC component), (5) NSC component PointSource vs.~S{\'e}rsic $\Delta$AIC values (explained in detail in section \ref{subsec:hstfits}), (6) quality of the NSC fits, see Section~\ref{subsec:qualitymodel}, (7) brief notes on the NSC fits.}
\end{deluxetable*}

We inspect the residuals of the centers for any visible pattern that is not being modeled.  If the central light excess appears to be resolved, we then replace the PointSource function with a S{\'e}rsic function. We consider this central component to be unresolved if the (i) $\Delta$AIC between the PointSource and S{\'e}rsic function is  $<$ 3000 and (ii) the difference in the surface brightness profile residuals between the PointSource and S{\'e}rsic function fits is $<$ 5$\%$.  We consider resolved sources to be NSCs.  Table \ref{tab:nscdata} indicates sources that are unresolved as well as the PointSource vs.~S{\'e}rsic $\Delta$AIC values. For some galaxies with known strong AGN, we use an additional PointSource component along with the NSC S{\'e}rsic component.  The evidence for AGN and details on the fits are given in Appendix~\ref{app:galnotes}. Out of the 33 galaxies we fit, we find four of them to have unresolved nuclear components (using the $\Delta$AIC $< 3000$ constraint discussed above).  We discuss these four sources in more detail in Section~\ref{subsec:nucfrac} below.

An example of this fitting procedure can be seen in the lower panel of Figure \ref{fig:modelingprcoess}, which shows the HST F814W image, the best-fitting F814W model image and the residual ratio images for three models: the first with no central (NSC) component, the second with a central PointSource component and the the third with the best-fitting NSC Sersıc component. From these three residual ratio images, we can visually judge the presence of a NSC component. The superiority of the third model, with the absence of any residual nuclear excess is evident. The AIC values for each of the 3 models are given below their respective residual ratio images; the NSC-S{\'e}rsic model has AIC $\sim -32,000$.

In most cases, we use the F814W image as the primary filter for modeling the NSCs. Once we have a good fitting model in the F814W filter, we then translate all the components to the F475W image as well as the F160W image -- we fit only for the intensity of these components, and keep the shape parameters fixed to scaled best-fit F814W values.  This ensures that we measure accurate colors for the NSCs (and other components). Allowing the NSC shape parameters to be fit independently in each filter can result in unphysical colors when e.g.~the NSC component is much larger in one filter than another. 

For galaxies that are very dusty in their centers (10 out of the 33), we instead use the F160W image as the primary band to fit our NSCs, and perform the methods outlined above on the F160W image directly, without translating to F814W.  The lower extinction in F160W helps us better fit the nuclear structure despite the lower spatial resolution.  In these cases, we still fit the F475W and F814W images with shape parameters fixed to the best-fit F160W values in order to derive colors for the NSCs in all three filters.   Once we have the best-fitting model in all three filters, we then calculate the magnitudes for all of the components -- explained in detail in section \ref{subsec:magcolor}.  Table~\ref{tab:nscdata} includes the primary filter used for modelling the NSCs.  

The top panel in figure~\ref{fig:sbprofile} shows an example of the 1-D surface brightness profiles we create to inspect the quality of the fit. These profiles were derived using circular aperture photometry with python's \texttt{photutils} library.  We ignore masked pixels in deriving the surface brightness of both the original image (green points) as well as the best-fitting model (blue solid line). The NSC S{\'e}rsic component (purple solid line) can be identified as the visible bump seen at the smallest radii. The bottom panel shows the residuals (model -- data) in magnitudes; radii where the data is brighter than the model have positive values in this panel. We also show the residuals for models with no central component (shown in red) and a central PointSource component (shown in orange). Here, we can observe changes in the residuals due to improving the NSC model fits. The residual based on the best-fitting model has residual below 0.05 magnitudes all the way out to 10\arcsec.

\subsection{Fit Quality, AGN, and Exceptions}
\label{subsec:qualitymodel}
Here we discuss the process of evaluating the goodness of the fits to the galaxy data. As mentioned in section \ref{subsec:iracfits} and section \ref{subsec:hstfits}, our quality check and criteria for adding new components for the each fit to the galaxy are based on (1) improvements in the residuals and (2) improvements in the AIC. 

Some of the galaxies in our NSC sample (especially in the MW-like subsample) have literature data that indicates they have an AGN. 
Detailed descriptions of these AGN components and associated X-ray sources are provided in Appendix~\ref{app:agnnotes}). For galaxies with a known AGN, we test whether an additional point source component needs to be included in the model as well, but find no cases where both a point-source and resolved NSC component provide a significantly improved fit.  In one case, NGC~1566, the AGN is bright enough (and in fact saturation artifacts affect the $r \lesssim 0.3\arcsec$ central region in the optical images) that no clear NSC component is visible. Accordingly, we do not present NGC~1566 in our NSC parameters table. 

Dust can also prevent us from obtaining good fits to the NSCs. In two galaxies (NGC 289 and NGC 7177) we are unable to determine accurate NSC morphologies due to dust obscuring the centers of all three bands. 

We rate the overall quality of our fits using a single number.  The quality values are as follows:\\
{\bf Quality 0:} These galaxies are those for which we do not obtain good fits for the NSCs -- this includes the galaxies discussed above, NGC~289, NGC~1566 and NGC~7177.  No fits to these galaxies are shown in Table~\ref{tab:nscprops}.\\
{\bf Quality 1:} Unresolved NSC components in four galaxies.  When comparing a point source component to a S{\'e}rsic component fit the $\Delta$AIC between these fits is $\lesssim$ 1000, giving minimal evidence that the nuclear component is resolved.  The nature of these sources is discussed in Section~\ref{subsec:nucfrac}.  \\
{\bf Quality 2:} dusty galaxies where dust absorption is evident all the way to the center.  However, we are able to estimate NSC properties using the F160W images in these galaxies.\\
{\bf Quality 3:} galaxies with complex central regions with poorly modeled structures (e.g., spiral arms) that result in large radial surface-brightness profile residuals ($>$0.1~mag).\\
{\bf Quality 4:} galaxies with some dust in the nuclear regions (but not crossing the center).   Fits have surface brightness profile residuals $<$0.1 magnitudes.\\
{\bf Quality 5:} galaxies with no dust within the central 2$\arcsec$.  Fits have surface brightness profile residuals $<$0.1 magnitudes.\\

There are 27 galaxies with quality $\geq$2.  For the plots in the paper, we exclude all NSCs with Quality $<$ 2, while Quality 2 fits are shown with open symbols.

\begin{figure}[!b]
\includegraphics[trim=0.2cm 2cm 2cm 3cm, clip, scale = 0.42]{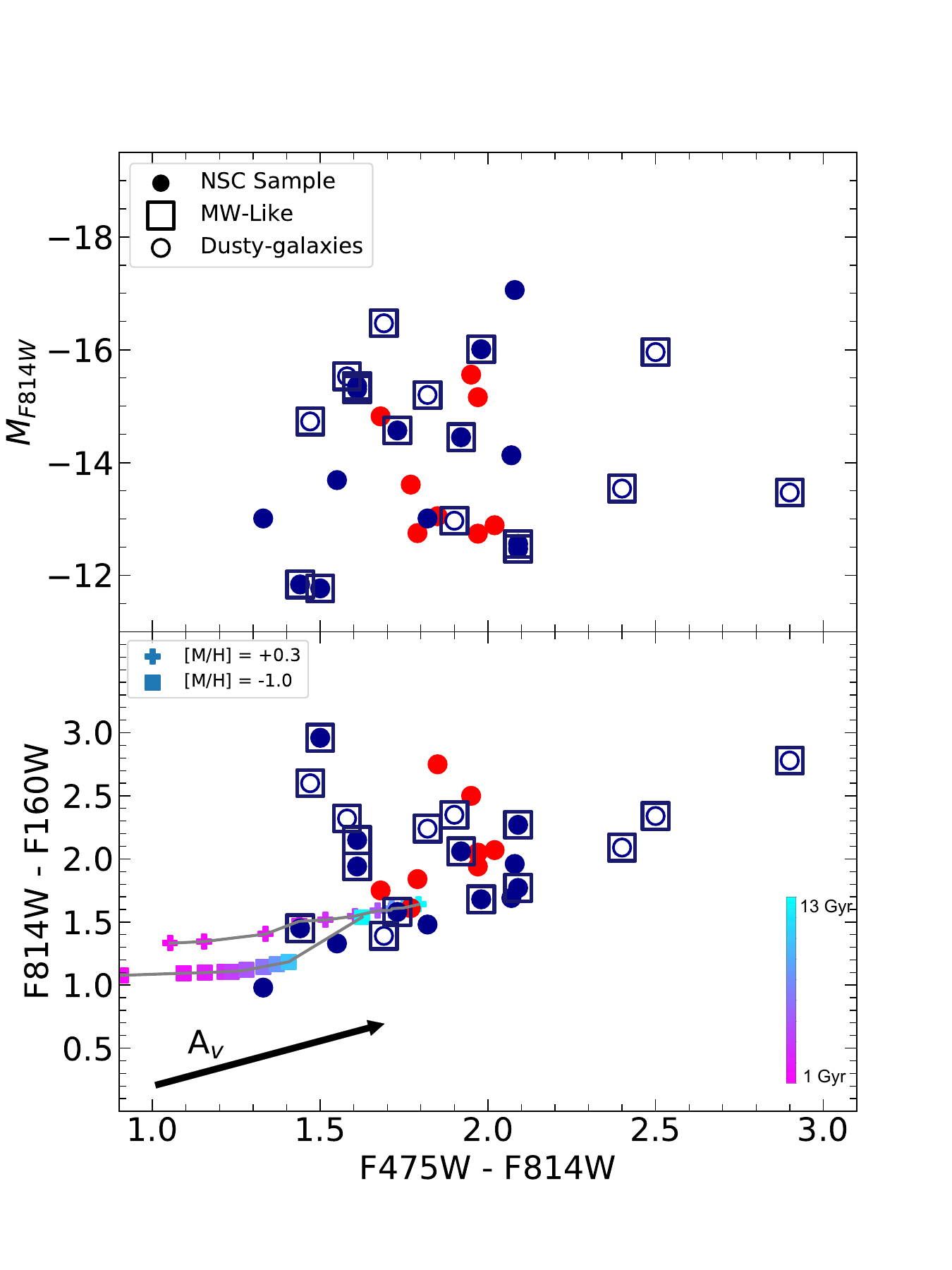}
\caption{\textbf{Top panel:} CMD with the NSC UVIS color VS F814W band absolute magnitude. The circles denote the NSCs in our sample. The MW-like subsample is denoted by the squares. The open circles represent the dusty galaxies with a NSC quality fit = 2.
\textbf{Bottom panel:} NSC color--color plot (UVIS vs. IR). Colored pluses and squares show PARSEC~1.2S SSP models \citep{bressan2012} at two metallicities (-1.0 and +0.3) and with ages from 1--13~Gyr (indicated by the color).  An extinction vector of corresponding to $A_V = 1$ is also shown.  All plotted magnitudes in both panels are foreground extinction corrected.}
\label{fig:colorplots}
\end{figure}

\begin{figure*}[!t]
\centering
\includegraphics[trim=0cm 0.5cm 0cm 0cm, clip, scale =0.55]{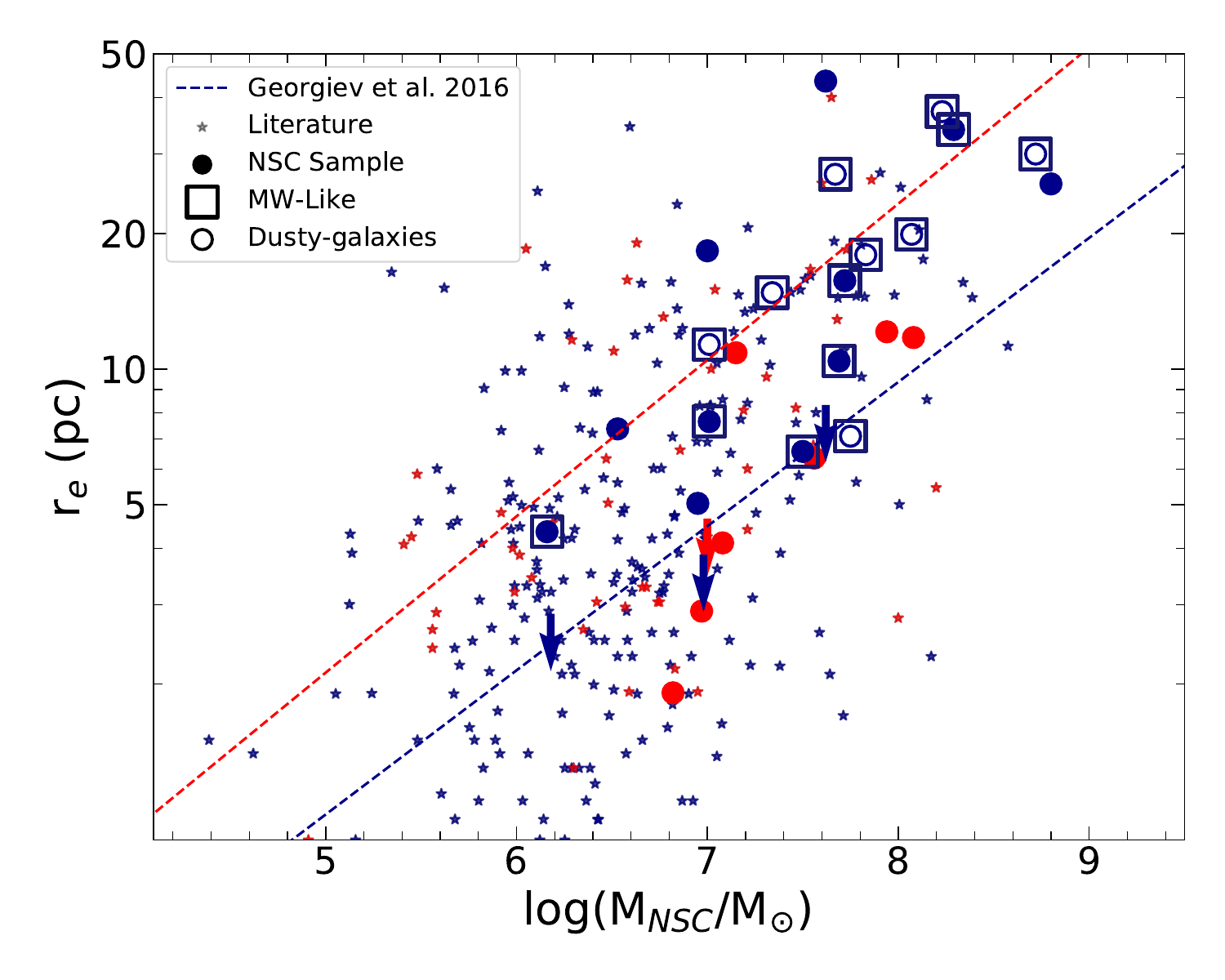}
\caption{The mass-size relation of our NSCs compared to previous literature measurements.  The circles denote our NSC sample galaxies, with surrounding squares indicating the MW-like subsample. Open circles indicate uncertain measurements due to dust (i.e., quality flag = 2 in Table~\ref{tab:nscprops}). The points are colored red or blue based on their Hubble type (early- and late- type respectively). The literature sample of NSCs from early- (red stars) and late- (blue stars) type galaxies are obtained from
\citet[][]{cote2006, spengler2017, georgiev2014, sanchez2019, erwin2012, eigenthaler2018, pechetti2020}.}
\label{fig:nscmassradius}
\end{figure*}

\subsection{Error Estimation}
\label{subsec:errors}
We estimate the errors on our \textsc{imfit} models using bootstrap resampling. This method allows us to capture asymmetric errors and the covariance of NSC parameters with the other fitted components. It also  provides more accurate errors than those estimated from the Leavenberg-Marquardt algorithm.~We performed 200 iterations of bootstrapping for each galaxy model. For the NSC models, the effective radii have median errors of $\sim$9\%, while the median error on S{\'e}rsic indices are $\sim$14\%. For ellipticities above 0.05, the errors on the ellipticity are just $\sim$7\%.

\section{NSC properties and results}
\label{sec:nscprops}

\subsection{Nucleation Fraction}
\label{subsec:nucfrac}
Of the 33 galaxies in our sample, we find unambiguous, resolved NSCs with radii $<$50~pc in 26.  Previously measured NSCs have typical effective radii of $\sim$3~pc with a small tail towards larger sizes and a cutoff suggested at $\sim$50~pc by \citet{neumayer2020}. In three galaxies, the presence of a nuclear cluster can't be constrained due to dust opaque enough to obscure the nucleus even in F160W (NGC~289 and NGC~7177), or the presence of a very bright AGN (NGC~1566) as discussed in Section~\ref{subsec:qualitymodel}.  In another four galaxies, the potential NSC components are unresolved: NGC 613 and NGC 1300 have a low luminosity AGN, and thus may not be NSCs. On the other hand in NGC 2775 and NGC 1440 the nuclear light appears to be stellar; specifically, the nuclear sources in both have very similar colors to the surrounding galaxy light, and the nuclear spectra \citep{ho1995} shows very little emission.  Therefore it is likely these two galaxies host compact NSCs that we just cannot resolve with HST.  The best-fit S\'ersic $r_e$ are used here as upper limits; for NGC~2775 and NGC~1440 these are 3.87~pc and 4.65~pc or 0$\farcs$042 and 0$\farcs$048.  Our ability to resolve NSCs in our CBS galaxies is complicated due to varying galaxy backgrounds, however, it does extend to more compact sources than these in some galaxies -- the most compact clearly resolved NSC is in NGC~4377 with an $r_e$ of 1.91~pc or 0$\farcs$22; this is similar to the limit on resolving NSCs found for galaxies at similar distances by \citet{cote2006}.

Using only the unambiguous resolved NSCs, we get a nucleation fraction of 78.8$_{-7.9}^{+6.2}$\% (26/33) for the full sample, with errors calculated using the Wilson interval.   The nucleation fraction is 68.4$_{-11.4}^{+9.5}$\% (13/19) for the MW-like subsample. On the other hand, we cannot exclude the presence of NSCs in any of our galaxies, thus the nucleation fraction in both samples could be as high as 100\%.  We note that our galaxy sample are all high mass ($\log \Mstar) > 10.1$) and mostly late-type galaxies; if we take all late-types in our sample, we get 76.0$_{-9.4}^{+7.4}$\% (19/25) with NSCs.

Two previous measurements exist for the nucleation fraction of massive late-type galaxies.  We took the data from \citet{neumayer2020} and \citet{hoyer2021} to find a comparable nucleation fraction for galaxies with ($\log \Mstar) > 10.1$); from the \citet{neumayer2020} compilation, 10/21 (47.6$_{-10.5}^{+10.7}$\%) galaxies in this mass range are nucleated, while in the \citet{hoyer2021} Local Volume sample, 10/15 (66.7$_{-13.8}^{+10.7}$\%) galaxies with nucleation measurements are nucleated.
Thus we find a higher nucleation fraction than either study, with our results being consistent with the measurement in \citet{hoyer2021}. Our values are much higher than the nucleation fraction of $\sim$30\% seen in early-type galaxies with similar mass \citep{neumayer2020,hoyer2021}.

\subsection{Magnitude and color}
\label{subsec:magcolor}
For all the NSCs, we estimate the magnitude using the \textsc{imfit} \texttt{makeimage} program, which can calculate fluxes and magnitudes for each component in the model. To determine the magnitudes of the NSC in the 3 HST bands, we divide the total counts by the total exposure time; 600s for F160W band, 500s for F814W band and 700s for F475W band. We then use the following zeropoints: 24.6949 in F160W, 24.684 in F814W, and 25.801 in F475W \footnote{\url{https://www.stsci.edu/hst/instrumentation/wfc3/data-analysis/photometric-calibration/uvis-photometric-calibration}}. 
These are Vega based zeropoints, and thus all magnitudes listed here are in the Vega system. We correct for foreground extinction using the $A_{F814W}$ values and conversions to the other two bands in Table~\ref{tab:galdata} and the notes to that table.  The extinction corrected NSC magnitudes in each filter band are presented in Table \ref{tab:nscprops}. 

Figure \ref{fig:colorplots} shows the color-magnitude and color-color (UVIS-IR) diagrams of the NSCs.  Padova PARSEC 1.2S single-stellar population models \citep{bressan2012} with ages from 1-13 Gyr and at two metallicities are overploted.   A majority of the galaxies are consistent with these models with modest extinction up to ($A_V \lesssim 2$).  Of these, only two require populations younger than $\sim$10~Gyr -- due to the age-extinction degeneracy it is not possible to separately constrain the ages and extinctions using our colors.  Almost half of the galaxies fall redwards of these SSP models in the F814W$-$F160W color in a way that is inconsistent with the reddening vector (which assume $R_V = 3.1$), in some cases by $>$1~magnitude.  This offset in F814W$-$F160W color could be due to (1) a mismatch between the data and SSP models, or (2) issues with calculating the NSC magnitudes.  To investigate this last issue, we measured aperture photometry within the center 0$\farcs$5, and compared these aperture colors to the model NSC colors.  For clusters with bright, prominent NSCs, the aperture and integrated NSC values agreed.  However, for fainter NSCs that make up a smaller light fraction of the galaxy, we see a blueward offset of up to 0.5 mags for the aperture magnitudes relative to the NSC model magnitudes. This blueward offset can be explained by a combination of the lower encircled energy in F160W and bluer surrounding bulge components.  
Overall, this test suggests we are accurately measuring the NSC colors with our model magnitudes.  While some NSCs have clear evidence for significant dust absorption (i.e.~the open circles), others appear dust-free, suggesting that the PARSEC models may under-predict the F814W$-$F160W colors of NSCs.

\subsection{Mass}
\label{subsec:mass}
We determine the NSC masses using the F814W magnitude and color-$M/L$ relations from \citet{roediger2015}. Specifically, we first convert our F814W magnitude to the Sloan $i$ band and then convert our HST F475W$-$F814W color to $g-i$ using relations derived from PARSEC 1.2S stellar population models \citep{bressan2012}:
\begin{eqnarray*}
    i - \mathrm{F814W} & = & 0.099 \times ({\mathrm{F475W}-\mathrm{F814W}) + 0.404} \\
    g-i & = & 0.92304 \times ({\mathrm{F475W} - \mathrm{F814W}}) - 0.48565
\end{eqnarray*}
We then use the $g-i$ color vs $i$ band magnitude relation of Table~2 in \citep{roediger2015} to determine the $i$-band $M/L$ ratio. The resulting NSC masses are presented in Table \ref{tab:nscprops}.
For our NSCs the $\log (\Mnsc / \Msun)$ values range from 5.7 to 8.74 with a median of 7.16 for all our NSCs and 7.26 for the NSCs in the MW-like subsample.
We determine the errors for the derived NSC mass using the bootstrap sampling (see Section~\ref{subsec:errors}), recalculating the luminosities, colors, and derived $M/L$s for each sample. 

\subsection{NSC size, mass relations}
\label{subsec:nscsizerelation}
We use the derived color and mass to understand our NSCs and compare them with the available literature from both early- and late-type galaxies. 
In Figure~\ref{fig:nscmassradius} we plot the radius of the NSCs (in pc) versus their derived masses. The solid circles denote our NSC sample galaxies, with surrounding squares indicating the MW-like subsample.  All galaxies are color coded into early- (red) and late- (blue) type. The dashed line show the~relationships for early- and late-type galaxies from \citet{georgiev2016}.
As has been found in many previous studies \citep[e.g.][]{hopkins2010,norris2014,georgiev2016,neumayer2020}, 
the NSC radii correlate with their masses, i.e., massive NSCs have a larger radii.  Our NSC sample improves the available literature, especially for the massive late-type galaxies ($> 10^{10} \Msun$).  Overall, the masses and radii of these clusters agree well with the overall trend seen in previously published  data (shown as small stars in Fig.~\ref{fig:nscmassradius}). 
\citet{georgiev2016} fit mass-radii relations and find that the NSCs in the late-type galaxies (blue-dashed line) are roughly two times smaller than the NSCs in early-type galaxies (red dashed line). Our data does not seem to support this difference; in particular most of the late-type galaxies in our sample fall above the blue dashed line, while all of the early-type galaxies fall below the red-dashed line.  This weakens the previous literature findings that there is a difference in the mass-radius relationship in early and late-type galaxies. 

\begin{figure}[!t]
\includegraphics[trim=0cm 2cm 2cm 3cm, clip, scale = 0.45]{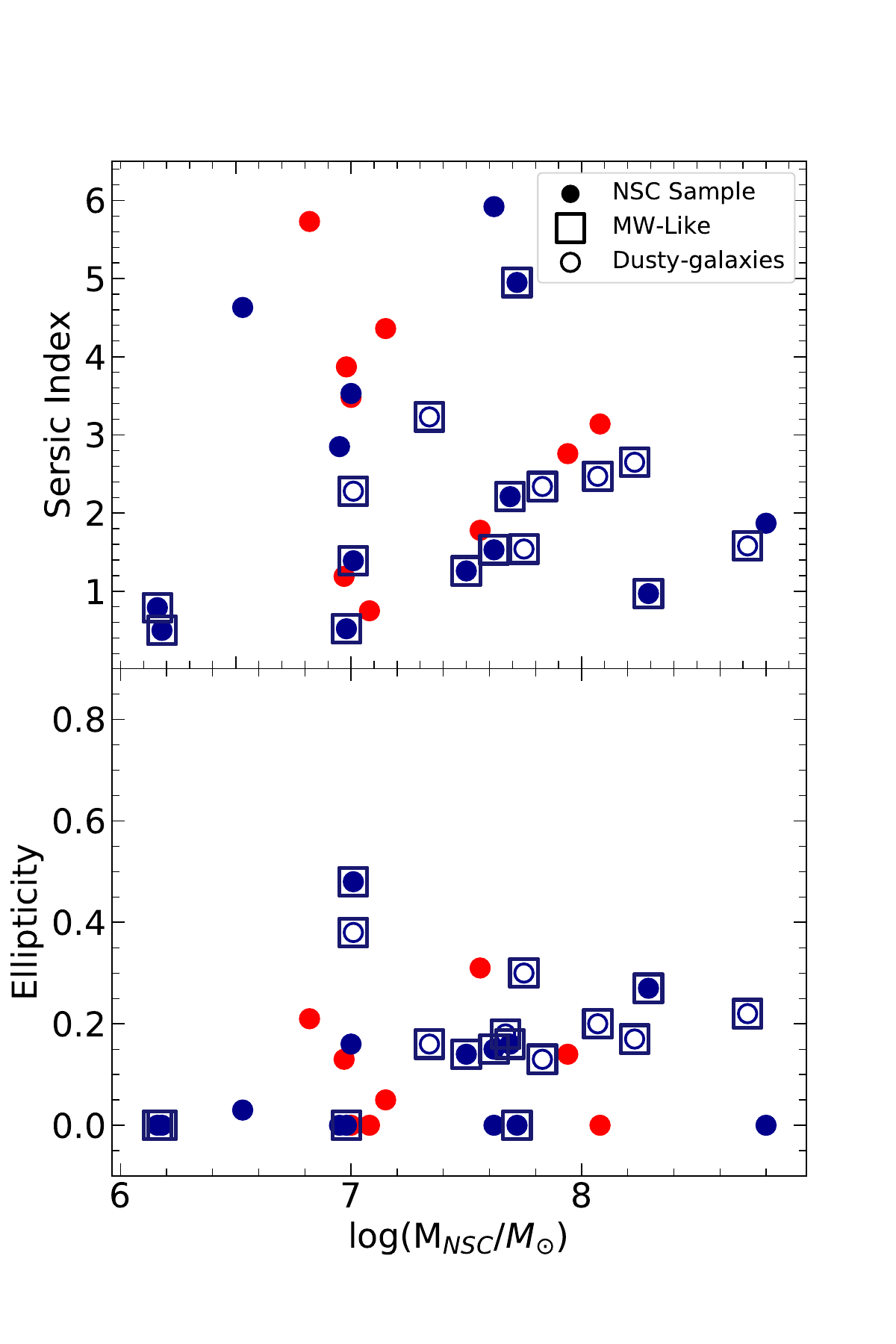}
\caption{\textbf{Top panel:} NSC mass vs. NSC S{\'e}rsic index. \textbf{Bottom panel:} NSC mass vs. NSC ellipticity. The solid circles denote our NSC sample galaxies, with squares indicating the MW-like subsample. All dusty galaxies (i.e., quality flag = 2) are denoted by open circles. The points are colored red or blue based on their Hubble type (Early- and Late- type respectively).}
\label{fig:nscmass_ell_n}
\end{figure}
The biggest NSCs ($> 25$ pc) in our sample are found in IC 2051, NGC 4380, NGC 4501, NGC 4699,  NGC 5248 and  NGC 7513. These largest objects may be ambiguous in their classification as NSCs, however, the continuity of the mass--radius relationship suggests these are related components. For NGC 7513, \citet{carollo2002} found the NSC to be a very compact source unlike in our model. The difference in the NSC model is discussed in detail in section \ref{subsec:nsclitcompare}.

Four objects (NGC~613, NGC 1300, NGC~1440 and NGC~2775) are unresolved in our sample but nonetheless appear to be NSCs; these are shown as upper limits on the mass radius diagram.  These galaxies fall on the compact side of the locus of previous NSC measurements and thus are significantly denser than typical NSCs.

\begin{deluxetable*}{ccccccccccc}[t!]
\centering
\tablecolumns{11}
\tablecaption{NSC S{\'e}rsic Parameters and derived properties.}
\label{tab:nscprops}
\tablehead{\colhead{Galaxy} & \colhead{PA$_{\rm nsc}$} & \colhead{$\epsilon_{\rm nsc}$} & \colhead{n$_{\rm nsc}$} & \colhead{r$_e$} & \colhead{m$_{\rm F475W}$} & \colhead{m$_{\rm F814W}$}& \colhead{m$_{\rm F160W}$} & \colhead{log(M$_{\rm NSC}$)} & \colhead{B/T$_{pm}$} & \colhead{L$_{bar}$}\\
\nocolhead{} & \nocolhead{} & \nocolhead{} & \nocolhead{} & \colhead{pc} & \colhead{mag} & \colhead{mag} & \colhead{mag} & \colhead{M$_{\rm \odot}$} & \nocolhead{} & \colhead{L$_{\rm \odot}$}\\
\colhead{(1)} & \colhead{(2)} & \colhead{(3)} & \colhead{(4)} & \colhead{(5)} & \colhead{(6)} & \colhead{(7)} & \colhead{(8)} & \colhead{(9)} & \colhead{(10)} & \colhead{(11)}}
\startdata
IC 2051 & 64.44$_{0.77}^{0.97}$ & 0.17$_{0.01}^{0.01}$ & 2.65$_{0.08}^{0.06}$ & 37.26$_{0.46}^{0.87}$ & 16.71$_{0.02}^{0.06}$ & 15.02$_{0.02}^{0.02}$ & 13.63$_{0.06}^{0.14}$ & 8.23$_{0.03}^{0.03}$ & 0.46 & 9.67\\
NGC 289 & -- & -- & -- & -- & -- & -- & -- & -- & -- & --\\
NGC 613 & 143.72$_{0.00}^{0.00}$ & 0.00$_{0.00}^{0.00}$ & 0.50$_{0.00}^{0.02}$ & 2.86$_{0.21}^{0.26}$ & 21.02$_{0.04}^{0.02}$ & 19.52$_{0.05}^{0.03}$ & 16.56$_{0.08}^{0.11}$ & 6.18$_{0.08}^{0.06}$ & 0.17 & 10.89\\
NGC 1097 & 59.14$_{1.24}^{3.08}$ & 0.16$_{0.01}^{0.01}$ & 3.23$_{2.35}^{2.20}$ & 14.77$_{0.37}^{0.17}$ & 17.60$_{0.07}^{0.07}$ & 16.12$_{0.02}^{0.04}$ & 13.53$_{0.03}^{0.27}$ & 7.34$_{0.11}^{0.09}$ & 0.16 & 10.94\\
NGC 1300 & 73.04$_{4.46}^{4.85}$ & 0.15$_{0.02}^{0.02}$ & 1.53$_{0.23}^{0.23}$ & 8.32$_{0.77}^{0.80}$ & 18.93$_{0.03}^{0.03}$ & 17.01$_{0.09}^{0.12}$ & 14.95$_{0.05}^{0.13}$ & 7.62$_{0.14}^{0.12}$ & -- & 10.80\\
NGC 1440 & 93.67$_{0.00}^{0.00}$ & 0.00$_{0.00}^{0.00}$ & 3.48$_{0.61}^{0.01}$ & 4.65$_{0.55}^{0.12}$ & 20.29$_{0.11}^{0.04}$ & 18.44$_{0.01}^{0.29}$ & 15.69$_{0.01}^{0.29}$ & 7.00$_{0.36}^{0.02}$ & 0.67 & 10.17\\
NGC 1566 & -- & -- & -- & -- & -- & -- & -- & -- & -- & --\\
NGC 2775 & 131.77$_{0.00}^{0.00}$ & 0.00$_{0.00}^{0.00}$ & 0.52$_{0.01}^{0.01}$ & 3.87$_{0.24}^{0.38}$ & 20.98$_{0.62}^{0.77}$ & 18.89$_{0.10}^{0.12}$ & 16.62$_{0.07}^{0.06}$ & 6.98$_{0.57}^{0.76}$ & 0.40 & --\\
NGC 3351 & 150.00$_{0.00}^{0.00}$ & 0.00$_{0.00}^{0.00}$ & 0.79$_{0.02}^{0.07}$ & 4.35$_{0.32}^{0.29}$ & 19.60$_{0.03}^{0.03}$ & 18.16$_{0.04}^{0.05}$ & 16.71$_{0.03}^{0.03}$ & 6.16$_{0.07}^{0.07}$ & 0.70 & 10.14\\
NGC 3368 & 60.68$_{1.24}^{0.98}$ & 0.16$_{0.04}^{0.00}$ & 2.21$_{0.71}^{1.05}$ & 10.41$_{1.51}^{1.00}$ & 16.42$_{0.10}^{0.66}$ & 14.81$_{0.05}^{0.07}$ & 12.87$_{0.56}^{0.06}$ & 7.69$_{0.25}^{0.21}$ & 0.75 & 10.67\\
NGC 3412 & 152.98$_{1.07}^{1.05}$ & 0.31$_{0.02}^{0.02}$ & 1.78$_{0.49}^{0.59}$ & 6.34$_{1.04}^{1.86}$ & 17.07$_{0.01}^{0.01}$ & 15.39$_{0.17}^{0.12}$ & 13.64$_{0.01}^{0.02}$ & 7.56$_{0.16}^{0.22}$ & 0.36 & 9.59\\
NGC 4237 & 116.78$_{2.21}^{4.07}$ & 0.03$_{0.01}^{0.02}$ & 4.63$_{0.33}^{0.03}$ & 7.36$_{0.05}^{0.63}$ & 19.70$_{0.17}^{0.01}$ & 18.37$_{0.05}^{0.16}$ & 17.39$_{0.09}^{0.04}$ & 6.53$_{0.14}^{0.08}$ & 0.05 & --\\
NGC 4321 & 55.85$_{3.23}^{4.45}$ & 0.13$_{0.05}^{0.15}$ & 2.34$_{0.21}^{0.32}$ & 17.92$_{1.57}^{1.24}$ & 17.53$_{0.03}^{0.02}$ & 15.71$_{0.07}^{0.05}$ & 13.47$_{0.06}^{0.06}$ & 7.83$_{0.07}^{0.07}$ & 0.14 & 10.63\\
NGC 4377 & 158.94$_{0.01}^{0.01}$ & 0.21$_{0.01}^{0.00}$ & 5.73$_{0.60}^{1.50}$ & 1.91$_{0.00}^{0.10}$ & 20.28$_{0.09}^{0.09}$ & 18.49$_{0.17}^{0.01}$ & 16.65$_{0.04}^{0.03}$ & 6.82$_{0.09}^{0.16}$ & 0.42 & 9.75\\
NGC 4380 & 149.79$_{1.46}^{1.06}$ & 0.18$_{0.02}^{0.01}$ & 6.95$_{0.13}^{0.16}$ & 27.04$_{1.57}^{2.87}$ & 19.88$_{0.01}^{0.02}$ & 17.47$_{0.04}^{0.04}$ & 15.39$_{0.04}^{0.06}$ & 7.67$_{0.05}^{0.04}$ & 0.17 & 9.12\\
NGC 4450 & 2.69$_{0.00}^{0.00}$ & 0.00$_{0.00}^{0.00}$ & 4.95$_{0.11}^{0.09}$ & 15.70$_{0.32}^{1.27}$ & 17.36$_{0.06}^{0.01}$ & 15.75$_{0.04}^{0.02}$ & 13.60$_{0.07}^{0.03}$ & 7.72$_{0.08}^{0.08}$ & 0.49 & 10.55\\
NGC 4501 & 143.74$_{1.74}^{1.54}$ & 0.22$_{0.01}^{0.01}$ & 1.58$_{0.06}^{0.07}$ & 29.99$_{0.34}^{0.34}$ & 17.63$_{0.01}^{0.05}$ & 15.13$_{0.06}^{0.04}$ & 12.79$_{0.07}^{0.07}$ & 8.72$_{0.03}^{0.07}$ & 0.14 & --\\
NGC 4531 & 6.89$_{5.47}^{4.79}$ & 0.05$_{0.01}^{0.01}$ & 4.36$_{0.34}^{0.29}$ & 10.86$_{1.50}^{0.80}$ & 19.07$_{0.03}^{0.01}$ & 17.30$_{0.05}^{0.02}$ & 15.69$_{0.17}^{0.09}$ & 7.15$_{0.03}^{0.02}$ & 0.22 & --\\
NGC 4548 & 147.12$_{0.00}^{0.00}$ & 0.20$_{0.03}^{0.04}$ & 2.47$_{0.38}^{0.44}$ & 19.89$_{2.78}^{2.83}$ & 20.49$_{0.12}^{0.17}$ & 17.58$_{0.02}^{0.03}$ & 14.80$_{0.04}^{0.03}$ & 8.07$_{0.11}^{0.14}$ & 0.36 & 10.31\\
NGC 4578 & 10.96$_{1.31}^{1.24}$ & 0.14$_{0.01}^{0.00}$ & 2.76$_{0.13}^{0.15}$ & 12.09$_{0.50}^{0.72}$ & 17.87$_{0.01}^{0.01}$ & 15.91$_{0.05}^{0.04}$ & 13.97$_{0.01}^{0.01}$ & 7.94$_{0.06}^{0.06}$ & 0.62 & --\\
NGC 4579 & 64.28$_{1.63}^{1.94}$ & 0.30$_{0.11}^{0.12}$ & 1.54$_{0.45}^{0.23}$ & 7.09$_{1.34}^{1.17}$ & 17.57$_{0.08}^{0.10}$ & 15.99$_{0.06}^{0.06}$ & 13.67$_{0.03}^{0.03}$ & 7.75$_{0.11}^{0.10}$ & 0.30 & 10.48\\
NGC 4608 & 145.63$_{1.12}^{1.80}$ & 0.13$_{0.04}^{0.04}$ & 1.19$_{0.07}^{0.09}$ & 2.90$_{0.10}^{0.10}$ & 20.42$_{0.04}^{0.04}$ & 18.45$_{0.01}^{0.01}$ & 16.40$_{0.06}^{0.05}$ & 6.97$_{0.03}^{0.04}$ & 0.72 & 10.27\\
NGC 4612 & 117.04$_{0.00}^{2.73}$ & 0.02$_{0.02}^{0.02}$ & 3.14$_{0.08}^{0.21}$ & 11.74$_{0.43}^{0.60}$ & 17.58$_{0.04}^{0.04}$ & 15.63$_{0.07}^{0.07}$ & 13.14$_{0.07}^{0.11}$ & 8.08$_{0.09}^{0.10}$ & -- & --\\
NGC 4643 & 114.16$_{0.00}^{0.00}$ & 0.00$_{0.00}^{0.00}$ & 0.75$_{0.34}^{2.70}$ & 4.11$_{0.53}^{3.86}$ & 20.53$_{0.03}^{0.04}$ & 18.52$_{2.00}^{0.11}$ & 16.45$_{0.02}^{0.02}$ & 7.08$_{0.15}^{1.25}$ & 0.37 & 10.61\\
NGC 4689 & 160.36$_{0.00}^{0.00}$ & 0.00$_{0.00}^{0.00}$ & 2.85$_{0.15}^{0.19}$ & 5.03$_{0.15}^{0.13}$ & 20.03$_{0.03}^{0.06}$ & 18.21$_{0.08}^{0.06}$ & 16.73$_{0.07}^{0.11}$ & 6.95$_{0.04}^{0.08}$ & 0.06 & 9.28\\
NGC 4698 & 66.76$_{1.23}^{1.46}$ & 0.48$_{0.04}^{0.01}$ & 1.39$_{0.24}^{0.65}$ & 7.64$_{1.46}^{0.29}$ & 20.62$_{0.19}^{0.20}$ & 18.53$_{0.05}^{0.05}$ & 16.76$_{0.08}^{0.14}$ & 7.01$_{0.16}^{0.17}$ & 0.44 & --\\
NGC 4699 & 152.00$_{0.00}^{0.00}$ & 0.00$_{0.00}^{0.00}$ & 1.87$_{0.28}^{0.44}$ & 25.75$_{3.42}^{2.27}$ & 16.46$_{0.07}^{0.03}$ & 14.37$_{0.06}^{0.07}$ & 12.41$_{0.01}^{0.12}$ & 8.80$_{0.03}^{0.09}$ & 0.14 & 10.19\\
NGC 5121 & 5.41$_{0.59}^{1.63}$ & 0.14$_{0.03}^{0.04}$ & 1.26$_{0.14}^{0.14}$ & 6.56$_{0.35}^{0.35}$ & 18.51$_{0.03}^{0.03}$ & 16.78$_{0.09}^{0.08}$ & 15.19$_{0.02}^{0.03}$ & 7.50$_{0.11}^{0.11}$ & 0.34 & 9.67\\
NGC 5248 & 107.37$_{0.97}^{0.91}$ & 0.27$_{0.01}^{0.01}$ & 0.97$_{0.04}^{0.03}$ & 33.99$_{0.20}^{0.28}$ & 17.16$_{0.03}^{0.02}$ & 15.18$_{0.01}^{0.05}$ & 13.50$_{0.04}^{0.03}$ & 8.29$_{0.03}^{0.04}$ & -- & --\\
NGC 5364 & 159.09$_{0.00}^{0.00}$ & 0.16$_{0.02}^{0.02}$ & 3.53$_{0.39}^{0.22}$ & 18.29$_{4.74}^{0.58}$ & 19.18$_{0.01}^{0.02}$ & 17.63$_{0.03}^{0.15}$ & 16.30$_{0.02}^{0.02}$ & 7.00$_{0.18}^{0.04}$ & 0.26 & --\\
NGC 6744 & 14.44$_{1.64}^{2.00}$ & 0.38$_{0.02}^{0.02}$ & 2.28$_{0.20}^{0.29}$ & 11.33$_{0.83}^{1.18}$ & 18.75$_{0.01}^{0.06}$ & 16.85$_{0.05}^{0.07}$ & 14.50$_{0.04}^{0.09}$ & 7.01$_{0.04}^{0.05}$ & 0.15 & 9.99\\
NGC 7177 & -- & -- & -- & -- & -- & -- & -- & -- & -- & --\\
NGC 7513 & 74.66$_{0.00}^{0.00}$ & 0.00$_{0.00}^{0.00}$ & 5.92$_{0.20}^{0.13}$ & 43.54$_{1.35}^{2.42}$ & 19.42$_{0.03}^{0.01}$ & 17.35$_{0.16}^{0.02}$ & 15.66$_{0.08}^{0.05}$ & 7.60$_{0.19}^{0.08}$ & 0.12 & 9.94\\
\enddata
\tablecomments{(1) Galaxy name, (2) NSC position angle (PA), (3) ellipticity, (4) S{\'e}rsic index, (5) effective radius, (6) F475W magnitude, (7) F814W magnitude, (8) F160W magnitude, (9) logarithmic stellar mass estimated using the $M/L$ ratio (see Section~\ref{subsec:mass}, (10) Photometric bulge to total ratio of the galaxy (see Section~\ref{subsec:bulgecorrelation}), (11) bar luminosity determined using the bar component magnitudes. All magnitudes are corrected for Galactic extinction.}
\end{deluxetable*}

\subsection{NSC mass vs. ellipticity and S{\'e}rsic index}
\label{subsec:nscmassandparams}
In Figure \ref{fig:nscmass_ell_n}, we compare the NSC mass with the NSC S{\'e}rsic index (top panel) and ellipticity (bottom panel). 

We see no correlation between NSC masses and S{\'e}rsic indices, indicating that the NSCs have a wide range of concentrations \citep[see also][]{hoyer2023}. One of the possible reasons for a high S{\'e}rsic index ($> 5$) is the presence of multiple components within the NSC, as seen in previous studies of very nearby NSCs where these components can be resolved \citep[e.g.,][]{seth2006, seth2010, nguyen2018}. The S{\'e}rsic indexes of our NSCs range from 0.1--5.9 with a median value of 2.7. The median value of the S{\'e}rsic indices of the MW-like subsample is 2.2. We have no evidence for multiple component NSCs in our sample except for in NGC 4612, in which we obtain a two component fit. We only plot the average ellipticity obtained from the two component NSC in Figure~\ref{fig:nscmass_ell_n} and do not plot the S{\'e}rsic index of this NSC. More information about the NSC fit for this galaxy is provided in the Appendix~\ref{app:galnotes}.

A trend of higher ellipticities in higher mass NSCs was seen in early-type galaxies by \citet{spengler2017}, however, we find no correlation between the NSC mass and ellipticity (bottom panel in figure \ref{fig:nscmass_ell_n}). The ellipticities of all our NSCs range from 0--0.5 with a median value of 0.14 (0.16 for the MW-like subsample).

\begin{figure*}[!t]
\includegraphics[trim=3.5cm 1cm 2.5cm 2cm, clip, scale =0.42]{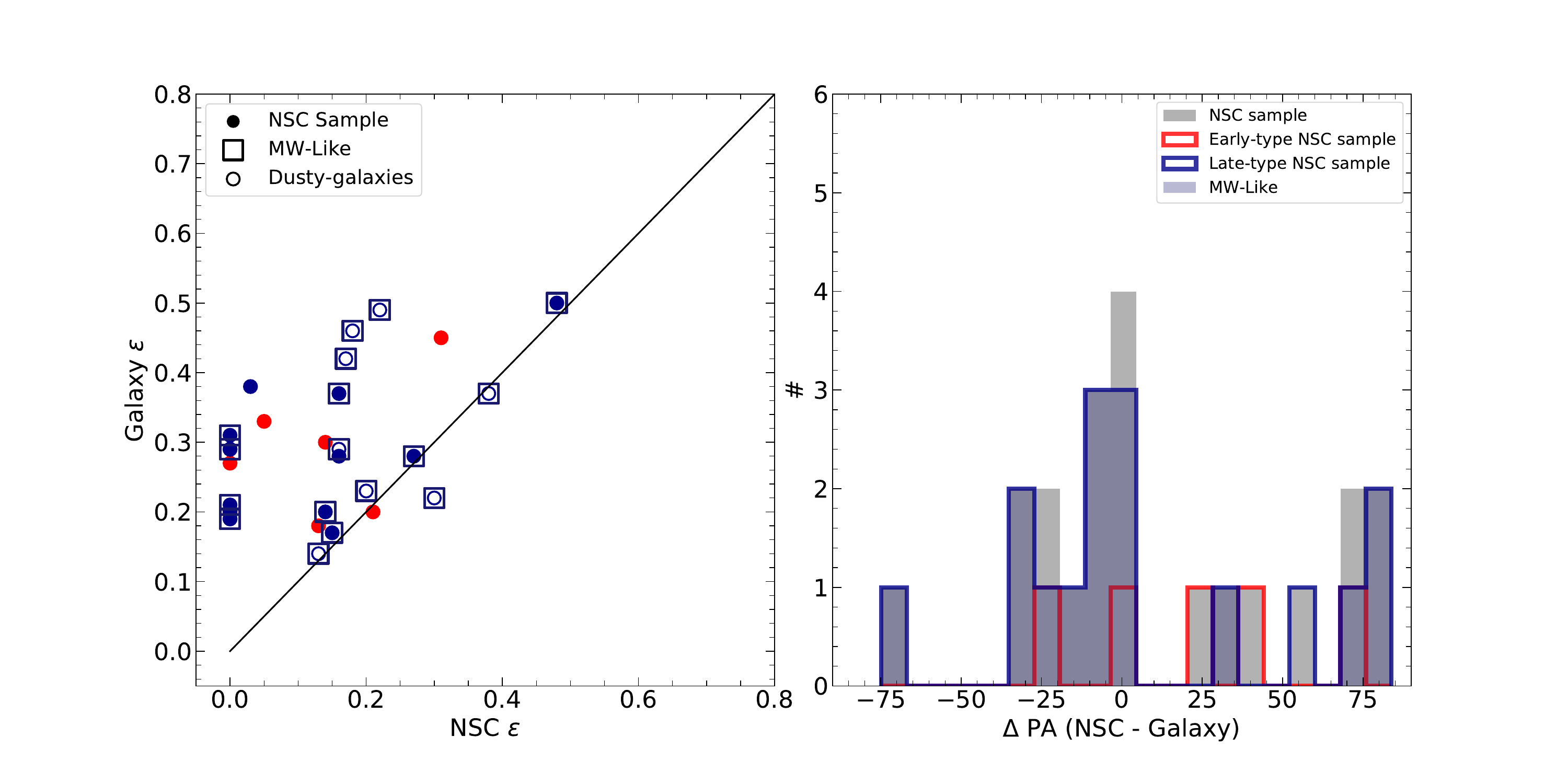}
\caption{\textbf{Left panel:} Galaxy ellipticity vs NSC ellipticity. The population of NSCs is typically rounder than their host galaxy disks, suggesting a less flattened distribution.  Circles denote the NSC sample with squares showing the MW-like subsample. The galaxies are colored red and blue based on their Hubble type (early and late type respectively). The NSCs with quality flag = 2, i.e., dusty centers are denoted by open circles. \textbf{Right panel:} Difference between NSC and Galaxy position angles ($\delta$PA). We plot the difference for all those NSCs whose $\epsilon$ $> 0.05$.  There is a clear preference for near-alignment between NSCs and their host galaxies' disks.}
\label{fig:galpa}
\end{figure*}

\subsection{NSC literature comparison}
\label{subsec:nsclitcompare}

The NSCs of five of the galaxies from our sample -- NGC 289, NGC 1566, NGC 4237, NGC 4612 and NGC 7513 -- have been studied previously in the literature. Two of these galaxies (NGC~289 and NGC~1566) have NSCs presented in \citet{carollo2002}; due to the bright AGN component (in NGC~1566) and dust (in NGC~289), we are unable to obtain reliable, unambiguous NSC fits for them, and so we focus on the other three objects, below.

\indent {\textbf{NGC 4237} --} This unbarred Virgo Cluster spiral galaxy has been previously studied in \citet[][]{georgiev2014}. They determine the effective radius of the NSC to be 0$\farcs$07 with an F814W magnitude of 17.74. They model the NSC using multiple images from HST including F606W and F814W band. From our work, the NSC in this galaxy has an effective radius of $0\farcs08$, with an F814W magnitude of 18.37.  So while the effective radii are similar, our fit to the NSC is considerably fainter than \citet{georgiev2014}.  This is likely due to our more careful modeling of the galaxy background.  We note we also are fitting higher resolution images than the wide field chip WFPC2 images fitted in \citet{georgiev2014}.
\\\\
\indent {\textbf{NGC 4612} --} This barred S0 Virgo Cluster galaxy was previously studied in \citet{cote2006} and \citet{spengler2017} as part of the ACS Virgo Cluster Survey. They determine the NSC in this galaxy to be unresolved with an effective radius of $0\farcs024$.  In our model, we find that the NSC is best fit by a two component model, a compact S\'ersic surrounded by a larger exponential with a combined effective radius of $0.14\arcsec$.  This model is preferred to a point source model with a $\Delta$~AIC of 5810. We can directly compare the F475W magnitudes of the sources; they find 17.73 (after conversion to Vega magnitudes), while our combined NSC has a magnitude of 17.66, thus these agree quite well.  We note that our approaches differ significantly; \citet{cote2006} fit 1-D profiles with a single S\'ersic background galaxy models, while we fit a more sophisticated galaxy model and fit in 2D.
\\
\indent {\textbf{NGC 7513} --}This barred galaxy was previously studied in \citet{carollo2002}. They modeled the NSC using NICMOS2 data in the F110W and F160W filters and determine the NSC in this galaxy to have an F160W magnitude of 18.3 and an effective radsius of  0$\farcs$06 (0.97 pc), slightly smaller than the 0$\farcs$075 pixels.   
We find a much bigger and brighter NSC component in this galaxy, with an effective radius of $0.45\arcsec$ (43.54 pc) and an F160W magnitude of 15.66; an unresolved point source or compact component provides a much worse fit to the nuclear regions.  The \citet{carollo2002} fits did not model the galaxy background at all, and we suspect that this methodological difference may be responsible for this discrepancy. 

\section{NSC-Galaxy relations}
\label{sec:nscgalrelations}

In this section, we discuss in detail the relation of the NSCs to the properties of their host galaxies. We also briefly discuss the relations between the NSC mass and supermassive black hole mass from \citet{saglia2016} for 4 galaxies in our sample in Section \ref{subsec:nscbhmass}.

\begin{figure*}[t!]
\centering
\includegraphics[trim=0cm 1cm 0cm 2.5cm, clip, scale =0.55]{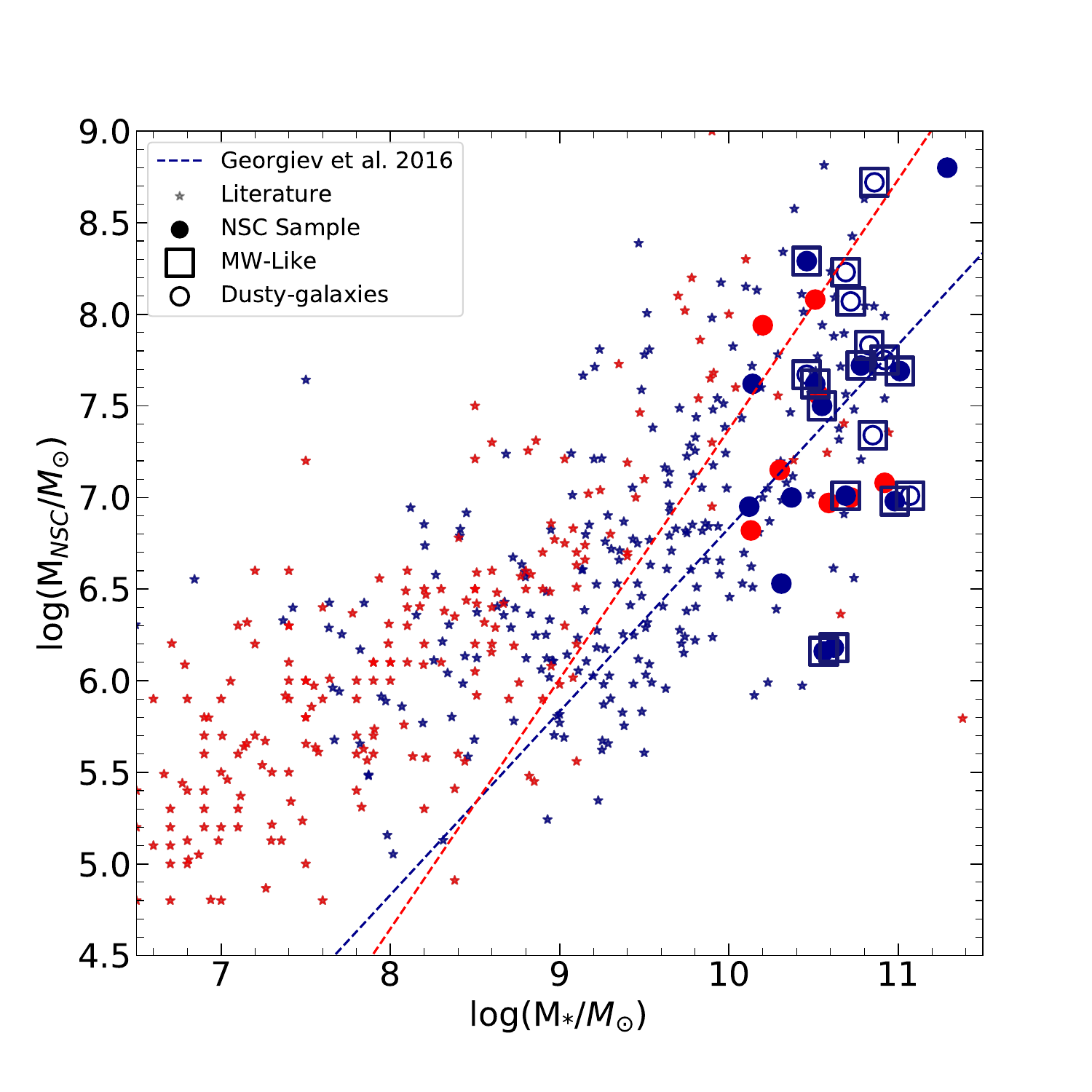}
\caption{The NSC mass is plotted against the host galaxy mass. The markings are the same as in Figure~\ref{fig:nscmassradius}. The literature sample contains both early- and late- type galaxies from \citet{cote2006, turner2012, georgiev2016, sanchez2019}.}
\label{fig:nscgal_mass}
\end{figure*}

\subsection{Structural Parameters}
\label{subsec:nscgalparam}
In Figure \ref{fig:galpa} we compare the ellipticity ($\varepsilon = b/a$) and position angle (PA) of our NSCs relative to their host galaxies. For the host galaxies, we use the PA and $\varepsilon$ from Table~\ref{tab:galdata}. It is important to note that the CBS selection criteria selects galaxies with inclinations between 35$^\circ$ and 60$^\circ$.  In the left panel of the figure, we see the NSCs have ellipticies equal to or smaller than the ellipticites of their host galaxies, suggesting that NSCs are typically rounder than their host galaxy disks. 

The right panel of the figure shows the difference in PA between the NSCs and the galaxies. Here we plot the difference for all NSCs with ellipticities $> 0.05$ where we can robustly estimate the NSC PA. We see that the distribution does not appear to be uniform as would be expected if there was no correlation between the galaxy disks and NSCs, but instead most of the NSCs have PAs within 25$^\circ$ of their host galaxies.  Using Kolmogorov-Smirnov and Anderson-Darling tests, we can reject the relative PAs being drawn from a uniform distribution at high significance ($p$-values of 0.0072 and 0.00075).  
This result is similar to what is seen in our Milky Way and in edge-on galaxies, where the NSC and galaxy PAs are typically aligned  \citep{seth2006,seth2008b,feldmeier2014}.  However, this correlation of PAs is not seen by \citet{georgiev2014}.  Their sample included a much wider range of inclinations, and less correlation between NSC and galaxy PAs would be expected to be visible in more face-on galaxies.  

The correlation of NSC and galaxy PAs suggests that NSCs are flattened and aligned with their large-scale galaxy disks.  This favors NSC formation from material in the disk -- either from gas accretion followed by {\em in situ} star formation, or by formation and inspiral of young star clusters \citep[e.g.][]{seth2006,agarwal2011,tsatsi2017}, and is consistent with strong rotation seen in many NSCs \citep{pinna2021}.  It disfavors NSC formation from inspiral of a more spherical distribution of globular clusters \citep[e.g.][]{tremaine1975,hartmann2011}.  This result is in agreement with previous work that suggests NSC formation is dominated by {\em in situ} star formation in more massive galaxies (log($M_\star/M_\odot$)$\gtrsim$9) like those in the CBS sample \citep{neumayer2020,fahrion2021,fahrion2022}.

\begin{figure}[!b]
\includegraphics[trim=0cm 2cm 2cm 3cm, clip, scale = 0.4]{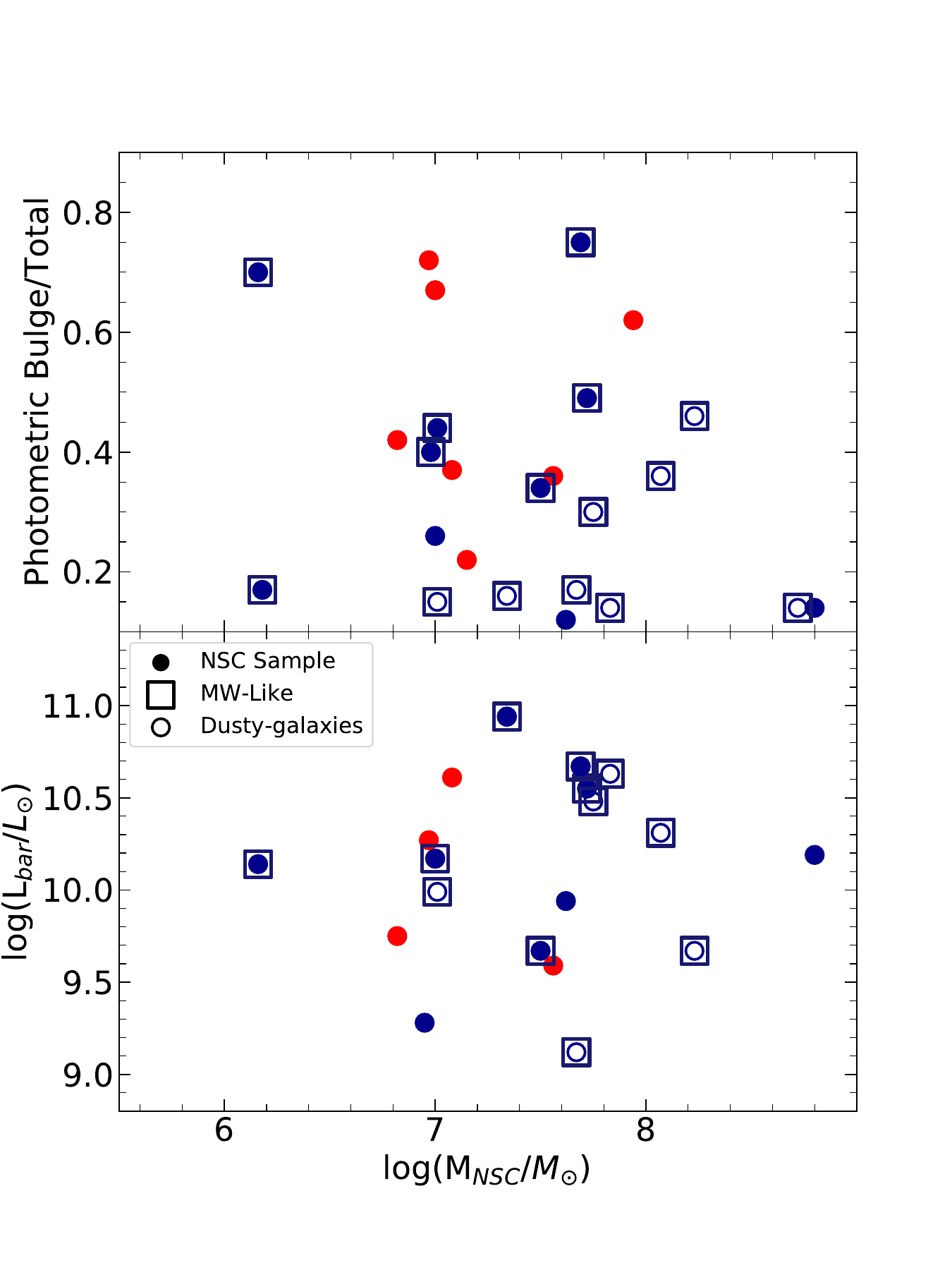}
\caption{The NSC mass plotted against~the~photometric~bulge to total ratio (top panel) and the bar luminosity (bottom panel). The markings are the same as in figure \ref{fig:colorplots}.}
\label{fig:btratio_bar}
\end{figure}

\begin{figure*}[ht!]
\includegraphics[trim=3.3cm 1cm 0cm 2cm, clip, scale =0.42]{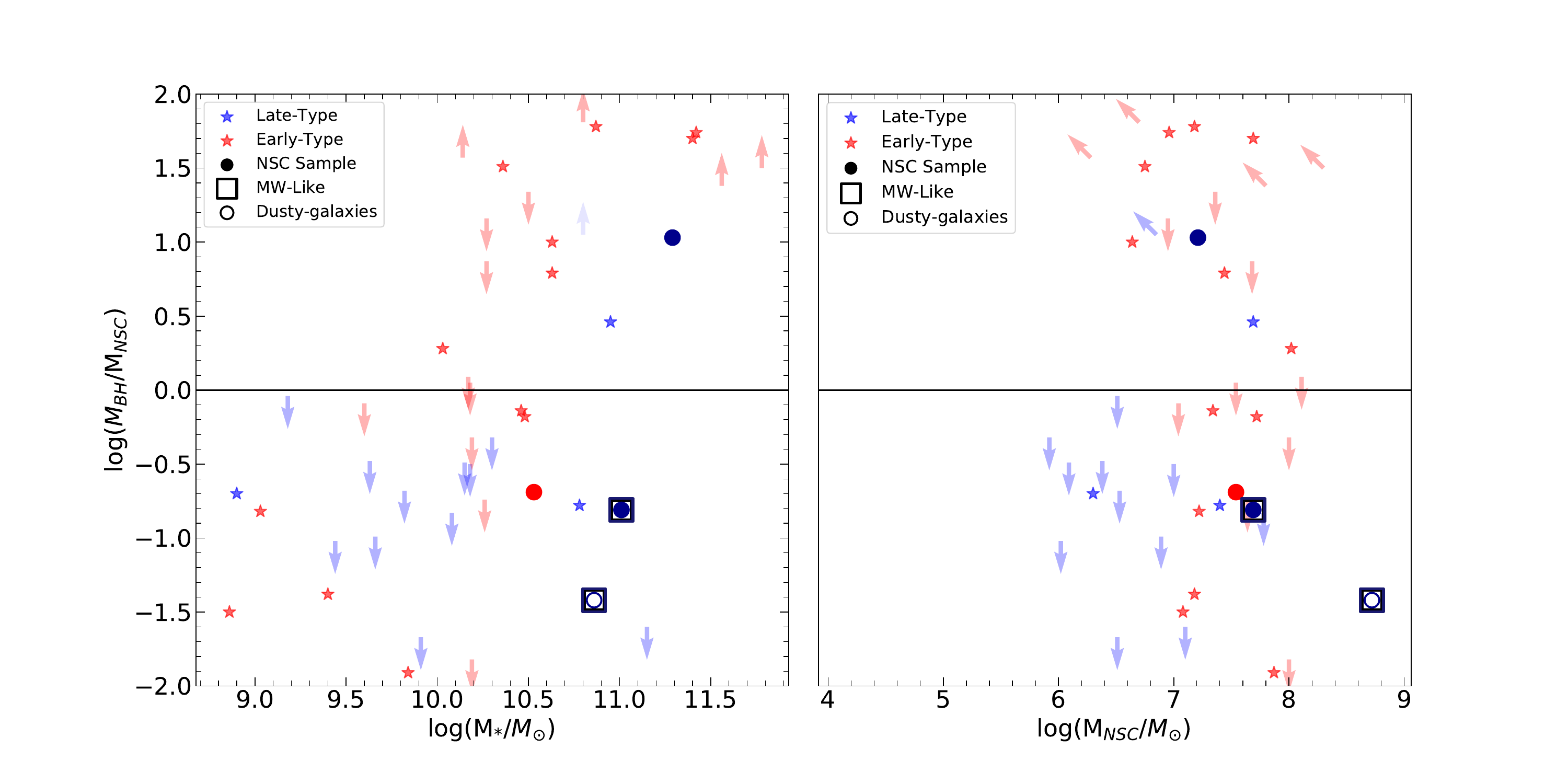}
\caption{The ratio of black hole mass to NSC mass is plotted against the host galaxy mass (\textbf{left panel}) and against the NSC mass (\textbf{right panel}) based on plots from \citet{neumayer2020} but adding in data with new NSC masses from our sample. The circles denote our NSC sample galaxies, with surrounding squares indicating the MW-like subsample. Open circles indicate uncertain measurements due to dust (i.e., quality flag = 2 in Table 3). The points are colored red or blue based on their Hubble type (Early- and Late- type respectively). Literature sample from Early- (red stars) and Late- (Blue stars) type galaxies are obtained from \citet[][]{seth2008, graham2009,neumayer2012,georgiev2016, nguyen2018}. Literature results with upper and lower limits for the BH and NSC masses are represented by the arrows. The solid horizontal line indicates equal NSC and BH masses in the galaxies.}
\label{fig:nscbh_mass}
\end{figure*}

\subsection{Correlations of NSC Mass with Host Galaxy Properties}
\label{subsec:nscgalmass}
Previous studies have found scaling relations between the mass of the NSC and its host galaxy properties.  This includes scaling relations with bulge luminosity, velocity dispersion, and total stellar mass \citep[e.g.,][]{ferrarese2006, wehner2006, rossa2006}. Initially, the NSC scaling relations were found to be similar to the BH scaling relations, but recent studies with more data over wider ranges of host galaxy properties have shown that, unlike BH scaling relations, the NSC mass correlates better with galaxy mass than bulge mass or stellar velocity dispersionf, and does not follow the same scaling relations  \citep[e.g.,][]{erwin2012,georgiev2016,sanchez2019}.

In Figure~\ref{fig:nscgal_mass}, we plot the NSCs mass against their host galaxy stellar mass. We observe that the bulk of our galaxies fall among the highest NSC masses as expected given their high galaxy stellar mass. All the NSCs in our sample have masses $> 10^6 \Msun$, with a median mass of $4.2 \times 10^{7} \Msun$.  The typical masses fall along the relation for late-type galaxies from \citet{georgiev2016}, with a tight cluster of points around the  median sample mass.  However there is also a very large, $>$2 order of magnitude spread in the NSC mass, with several significant low outliers including NGC 3351 ($\log \Mstar = 10.56$ and $\log \Mnsc = 6.16$) and NGC 6744 ($\log \Mstar = 11.07$ and $\log \Mnsc = 7.01$) and the very compact NSC in NGC 2775 ($\log \Mstar = 10.98$ and $\log \Mnsc = 6.98$). 
This broad range of masses suggests a wide range of formation and evolutionary processes in the NSCs in these massive (mostly late-type) galaxies. As noted in Section~\ref{subsec:nucfrac}, the nucleation fraction in our NSC sample is much higher than the nucleation fraction of early-type galaxies with similar mass \citep{cote2006,neumayer2020}, where binary BH mergers might have destroyed NSCs \citep[e.g.][]{milosavljevic2001}.  With the ongoing formation of NSCs in late-type galaxies \citep[e.g.][]{walcher2006,rossa2006}, the absence of an NSC after a binary BH merger would likely be short-lived and the existence of low mass outliers may trace galaxies where NSCs have been destroyed in the relatively recent past.  The compact radii ($< 5$ pc) of two of these clusters may be due to the reformation resulting in more compact NSCs; recent star formation is seen to be centrally concentrated in nearby NSCs including the Milky Way \citep{feldmeier-krause2015}, M31 \citep{lauer2012}, and other nearby late-type NSCs \citep{carson2015}.  

\subsubsection{Bulge and Bar Relations}
\label{subsec:bulgecorrelation}
Given that NSCs are located in the centers of galaxies, it is interesting to understand how they relate with the bulge and bar components in their hosts. Understanding this relationship might provide insights into their formation mechanism. In this section, we discuss in detail the correlations of the NSC mass with the fraction of light in the photometric bulge and the luminosity of the bar components hosted by our sample of galaxies. In Figure \ref{fig:btratio_bar}, we plot the NSC mass against the photometric bulge to total ($B/T$) light ratio (top panel) and the bar luminosities (bottom panel). We define the photometric bulge to consist of all the components (including the NSCs) except the bar and the disk. We determine the luminosity of the bar components in the galaxy, integrating the flux of the bar and the boxy-peanut bulge components (we note that 8/26 of the galaxies with NSCs lack bar components and are not included). Also, since we do not model the disk in NGC 1300 and NGC 5248, we exclude these galaxies from the top panel in the figure.

We find no correlation between the NSC mass and the photometric $B/T$ ratio in our sample. This indicates that the bulges in galaxies appear to be uncorrelated with the formation of the NSCs.  Similarly, we do not see any correlation between the NSC mass and the bar luminosities.

\subsection{NSC-Black Hole relations}
\label{subsec:nscbhmass}

NSCs and BH are found to co-exist in many massive galaxies \citep[$>$ 10$^{9} \Msun$, e.g.][]{seth2008,gonzalez-delgado2008}.  Early studies of scaling relations and the relative masses of these quantities suggested there may be a transition between NSCs dominating the nuclear mass in low mass galaxies and BHs dominating in higher mass galaxies \citep{ferrarese2006,wehner2006,graham2009}. The recent compilation of co-existing BH and NSC mass measurements in \citet{neumayer2020} shows a clear trend where NSCs are typically more massive than BHs in lower-mass galaxies, while the opposite is true in higher mass galaxies.  NSCs become less common in massive early-type galaxies ($> 10^{10} \Msun$). This could be due to the dynamical impact of binary BH mergers
\citep[e.g.][]{milosavljevic2001,antonini2015}. 
However, the trend in NSC/BH mass is not a simple one -- 
This large scatter can be seen in the very different relative masses of the BHs and NSCs in the Milky Way (where the NSC is $\sim$10$\times$ the BH mass) and M31 (where the opposite is true).  Unfortunately, a lack of NSC mass measurements in massive spiral galaxies has limited our ability to make this comparison more widely for MW-like galaxies.

From our galaxies with good NSC measurements, 4 galaxies have black hole mass measurements available in the \citet{saglia2016} compilation (NGC 3368, NGC 3412, NGC 4501, and NGC 4699). In Figure~\ref{fig:nscbh_mass} we show the ratio of BH and NSC mass of these galaxies against their host galaxy stellar mass (left panel) and the NSC mass (right panel) added to the data from \citet{neumayer2020}. 
The solid line in the figure represents equal NSC and BH mass. The objects above this line (including our measurement for the very massive late-type galaxy NGC~4699) have a more massive BH than NSC.  Our measurement confirm that there are a wide range of BH to NSC mass ratios in massive galaxies.  

\section{Conclusions}
\label{sec:conclusion}
In this paper, we have presented the photometric and morphological analyses of nuclear star clusters in 33 nearby galaxies from the Composite Bulges Survey. This includes a subsample of 20 MW-like galaxies with spiral morphologies (T = 1--4) and stellar mass from $10^{10.4} \Msun$ to $10^{11.1} \Msun$. This MW-like subsample is a complete volume-limited sample of galaxies similar to the Milky Way that also meet the distance ($< 20$ Mpc), inclination (35 to 60$^\circ$), and Galactic latitude ($|b| > 20^\circ$) criteria of the complete CBS sample.

Using \textsc{imfit}, we obtain accurate models for the nuclear regions of the galaxies.  We model the NSCs using S{\'e}rsic profiles in three HST filters and derive their sizes, colors, and masses. We present the S{\'e}rsic profile fit parameters of the NSCs in Table~\ref{tab:nscdata} and their derived properties in Table~\ref{tab:nscprops}. 

Our main results are:
\begin{itemize}
    \item We clearly identify NSCs in 78.8$_{-7.9}^{+6.2}$\%  of our 33 galaxies, and  68.4$_{-11.4}^{+9.5}$\% for the MW-like subsample. NSCs may be present in other galaxies, but are missed due to dust or AGN, thus these nucleation fractions are lower limits.  This work significantly expands the number of nucleated galaxies known in higher mass, late-type galaxies.  The nucleation fractions are higher than, but consistent within the errors of the determination in \citet{neumayer2020}.  
    \item We calculated the mass of our NSCs using color-$M/L$ relations, and find a median mass of NSCs in our galaxies of $\log (\Mnsc / \Msun) = 7.16$.  
    \item Our NSCs are consistent with the mass-radius relationship of literature NSCs (Figure.~\ref{fig:nscmassradius}).  They also follow the galaxy stellar mass-NSC mass relation for late-type galaxies from \citet{georgiev2016}, significantly expanding the sample of NSCs at the high mass end (Figure.~\ref{fig:nscgal_mass}).  
    \item We find a large scatter in NSC mass over a small range of galaxy mass, with two prominent low-mass outliers.  These outliers also have small radii, suggesting a possible difference in NSC formation mechanism or evolution.  
    \item Our NSCs are preferentially aligned with but are less flattened than their host galaxy disks. This alignment suggests these NSCs are forming either from gas accretion or star clusters inspiraling from the disk due to dynamical friction.
    \item Our NSCs do not show any correlation with the bar luminosity or photometric $B/T$ ratios. 
    \item We add four more galaxies to the small number of galaxies with known NSC and BH masses.  One has a BH $\sim 10$ times the mass of the NSC, while the others have NSCs that greatly exceed their BH masses.  This confirms that massive galaxies have a wide range of NSC-to-BH mass ratios.  
\end{itemize}

\textbf{ACKNOWLEDGEMENTS}\\
We would like to thank David Ohlson for sharing his galaxy catalog with us pre-publication. We would also like to thank Luis Ho for providing us with optical images from the Carnegie-Irvine Galaxy Survey and Preben Grosb{\o}l for providing us with near-IR images from \citet{grosbol2012}.

This research is based on observations made with the NASA/ESA \textit{Hubble Space Telescope} obtained from the Space Telescope Science Institute, which is operated by the Association of Universities for Research in Astronomy, Inc., under NASA contract NAS 5–26555. These observations are associated with program HST-GO-15133.  Support for this work was provided from that program.  J.M.A. acknowledges the support of the Viera y Clavijo Senior program funded by ACIISI and ULL. This work is based in part on observations made with the \textit{Spitzer} Space Telescope, which was operated by the Jet Propulsion Laboratory, California Institute of Technology under a contract with NASA.  It has also made use of the NASA/IPAC Extragalactic Database (NED), which is operated by the Jet Propulsion Laboratory, California Institute of Technology, under contract with the National Aeronautics and Space Administration.  We also made use of data from the Sloan Digital Sky Survey.

\software{Astropy \citep{Astropy,Astropy2}, DrizzlePac \citep{DrizzlePac}, Imfit \citep[][v1.8]{imfit}, grizli \citep{grizli}}


\bibliography{NSC_paper}
\bibliographystyle{aasjournal}
\clearpage

\appendix
\section{AGN detections in the NSC sample of galaxies}
\label{app:agnnotes}
\begin{deluxetable}{ccccl}[!h]
\label{tab:agntable}
\tablecaption{Nuclear Classification \& AGN Table}
\tablecolumns{5}
\tablehead{\colhead{Name} & \colhead{Nuclear Class} & \colhead{AGN Source} & \colhead{log($L_X$ [ergs/s])} & \colhead{Notes}}
\startdata
NGC0289 &  &  &  &  \\
NGC0613 & S? & 2 &  &  \\
NGC1097 & L/S1 & 2 & 40.96 & (a) \\
NGC1300 & T & 2 & 40.12 &  \\
NGC1440 &  &  &  &  \\
IC2051 &  &  &  &  \\
NGC1566 & S1.2 & 3 & 41.1-42.5  & (b) \\
NGC2775 &  &  &  &  \\
NGC3351 (M95) & H & 1 &  &  \\
NGC3368 (M96) & L2 & 1 &  &  \\
NGC3412 & A & 1 &  &  \\
NGC4643 & T2 & 1 &  &  \\
NGC4699 &  &  &  &  \\
NGC5121 &  &  &  &  \\
NGC5248 & H & 1 & 38.32  & \\
NGC5364 & H & 1 &  &  \\
NGC6744 &  &  &  &  \\
NGC7513 &  &  &  &  \\
NGC4237 &  &  &  &  \\
NGC4321 (M100) & T2 & 1 &  &  \\
NGC4377 &  &  &  &  \\
NGC4380 & H & 1 &  &  \\
NGC4450 & L1.9 & 1 & 40.55  & \\
NGC4501 (M88) & S2 & 1 & 40.16 &  \\
NGC4528 &  &  &  &  \\
NGC4531 &  &  &  &  \\
NGC4548 (M91) & L2 & 1 & 39.93  & \\
NGC4578 & A & 2 &  \\
NGC4579 (M58) & S1.9/L1.9 & 1 & 41.61 &  \\
NGC4608 &  &  &  &  \\
NGC4612 & A & 1 &  &  \\
NGC4689 & H & 1 &  &  \\
NGC4698 & S2 & 1 & 38.93 &  \\
\enddata
\tablecomments{Nuclear class gives nuclear classifications based on optical spectra: L=LINER, S=Seyfert A=Absorption, H=HII, and T=Transition spectrum, and number indicates Type 1 (broad line) or Type 2 (narrow line). AGN Source provides the reference of the AGN nuclear classification: (1) \citet{ho97}, (2) compiled by \citet{bi20}, (3) \citet{alloin85}.  The log($L_X$ [ergs/s]) give 2-10~keV X-ray luminosities; all are from  \citet{bi20} rescaled to the distances for each galaxy used in this paper except where noted.  Notes: (a) changing look AGN including change from LINER to Seyfert \citet{storchi-bergmann93}, classified as L1 in \citet{bi20}, (b) X-ray fluxes varying values from \citep{liu22}, changing look AGN \citep[e.g.][]{dasilva17}. }
\end{deluxetable}
\clearpage

\section{Common Components Used in Our \textsc{imfit} Modeling}
\label{app:components}

Below we list the most frequently used components used in our \textsc{imfit} fits.   

\begin{itemize}
\item FlatSky -- a uniform sky background (as discussed in section \ref{subsec:skysubtraction}).
\item Exponential -- an elliptical 2D exponential function. We use this primarily for fitting the disk and other highly elliptical components.
\item BrokenExponential -- two exponential zones having different scale lengths joined by a transition region of variable sharpness. We use this for fitting the outer disk component in some galaxies.
\item GaussianRing -- an elliptical ring with a radial profile consisting of a Gaussian function. We use this to ring or pseudoring features, such as nuclear rings.
\item FlatBar -- this is meant to represent the outer parts of bars, with a major-axis broken-exponential profile and a single-exponential minor-axis profile; it is described further in \citet{erwin2021}.
\item Sersic\_GenEllipse -- an elliptical 2D Sérsic function using generalized ellipses (``boxy'' to ``disky'' shapes). We use this to fit  boxy-shaped bulge features.
\item Sersic -- an elliptical 2D S{\'e}rsic function. We use this to fit NSC and the (non-boxy) bulge features.
\item PointSource -- a scaled representation of the image PSF, used primarily to model unresolved AGN emission.  
\end{itemize}

We provide detailed notes on each galaxy in the next section, including a handful of additional components other than these listed here. 

\section{Detailed notes on individual galaxy decomposition from the NSC sample}
\label{app:galnotes}
In this section, we describe the components of the best-fitting models for each galaxy in the NSC sample. The models are obtained in the primary band filter provided in Table~\ref{tab:nscdata} (column 3). For each component, we provide our best interpretation of what kind of structure it is, the \textsc{imfit} image function used (in brackets), and then the best-fit parameter values. The ordering generally reflects an inside-out description. Note that bar ``spurs'' refers to the outer, more elongated part of a bar, outside of the B/P-bulge part of the bar \citep[see, e.g.][]{erwin2013,erwin2021}. 
\vspace{6pt}
\\
{\bf \large IC 2051}
\begin{itemize}
    \item NSC [\textit{Sers\'ic}] (PA=17.0, $\epsilon$=0.18, n=2.65, $\mu_{e}$=14.60, $r_{e}$=0.39\arcsec)
    \item Nuclear disk(?) [\textit{Sers\'ic}] (PA=22.3, $\epsilon$=0.40, n=0.98, $\mu_{e}$=15.95, $r_{e}$=2.98\arcsec)
    \item B/P bulge [\textit{Sers\'ic$\_$GenEllipse}] (PA=27.4, $\epsilon$=0.40, n=1.44, $\mu_{e}$=18.22, $r_{e}$=7.06\arcsec)
    \item Bar spurs [\textit{FlatBar}] (PA=144.7, $\epsilon$=0.63, $\mu_{0}$=17.11, R$_{brk}$=19.6\arcsec)
    \item Inner ring [\textit{GaussianRingAZ}] (PA=12.0, $\epsilon$=0.29, A$_{maj}$=19.15, A$_{min-rel}$=27.34, $R_{ring}$=28.10\arcsec, $\sigma_{r}$=6.44\arcsec)
    \item Outer disk [\textit{BrokenExponential}] (PA=23.3, $\epsilon$=0.45, $\mu_{0}$=19.15, R$_{brk}$=60.17\arcsec)
\end{itemize}
\vspace{3pt}
{\bf \large NGC 289}\\
We do not trust the NSC measurements for this galaxy due to strong dust lanes obscuring the NSC, as mentioned in Section~\ref{subsec:qualitymodel}. Hence, we do not provide the best-fit models for this galaxy. \\
\\
{\bf \large NGC 613}\\
The NSC in this galaxy is unresolved (see section~\ref{subsec:nucfrac}). For the purpose of determining the NSC properties, we provide a Sers\'ic component fit. Note that the FlatBar component is offset by $\sim 1\arcsec$ from the other components.
\begin{itemize}
    \item NSC [\textit{Sers\'ic}] (PA=0, $\epsilon$=0.0, n=0.5, $\mu_{e}$=11.53, $r_{e}$=0.03\arcsec)
    \item Classical bulge(?) [\textit{Sers\'ic}] (PA=144.3, $\epsilon$=0.11, n=0.79, $\mu_{e}$=14.88, $r_{e}$=0.71\arcsec)
    \item Nuclear disk [\textit{Exponential}] (PA=151, $\epsilon$=0.32, $\mu_{0}$=13.94, h=3.23\arcsec)
    \item B/P bulge [\textit{Sers\'ic$\_$GenEllipse}] (PA=160, $\epsilon$=0.32, n=0.2, $\mu_{e}$=18.90, $r_{e}$=16.4\arcsec)
    \item Bar spurs [\textit{FlatBar}] (PA=162, $\epsilon$=0.92, $\mu_{0}$=17.70, R$_{brk}$=51.64\arcsec)
    \item Outer disk [\textit{Exponential}] (PA=154, $\epsilon$=0.41, $\mu_{0}$=17.93, h=47.06\arcsec)
\end{itemize}
\vspace{3pt}
{\bf \large NGC 1097}
\begin{itemize}
    \item NSC [\textit{Sers\'ic}] (PA=117.1, $\epsilon$=0.16, n=3.23, $\mu_{e}$=13.20, $r_{e}$=0.21\arcsec)
    \item Inner-bar B/P bulge [\textit{Sers\'ic}] (PA=178.1, $\epsilon$=0.05, n=1.09, $\mu_{e}$=15.32, $r_{e}$=3.08\arcsec)
    \item Inner-bar spurs [\textit{FlatBar}] (PA=86.3, $\epsilon$=0.62, $\mu_{0}$=15.00, R$_{brk}$=6.39\arcsec
    \item Nuclear ring [\textit{GaussianRing}] (PA=29.7, $\epsilon$=0.18, A=15.91, $R_{ring}$=10.19\arcsec, $\sigma_{r}$=0.89\arcsec)
    \item Nuclear disk [\textit{BrokenExponential}] (PA=10.4, $\epsilon$=0.20, $\mu_{0}$=16.61, R$_{brk}$=12.93\arcsec)
    \item Outer-bar B/P bulge [\textit{Sers\'ic}] (PA=11.1, $\epsilon$=0.29, n=0.38, $\mu_{e}$=18.21, $r_{e}$=27.72\arcsec)
    \item Outer-bar spurs [\textit{FlatBar}] (PA=26.3, $\epsilon$=0.92, $\mu_{0}$=17.75, R$_{brk}$=73.74\arcsec)
\end{itemize}
{\bf \large NGC 1300}
\begin{itemize}
    \item NSC [\textit{Sers\'ic}] (PA=59.9, $\epsilon$=0.15, n=1.53, $\mu_{e}$=14.49, $r_{e}$=0.09\arcsec)
    \item Classical bulge(?) [\textit{Sers\'ic}] (PA=90.3, $\epsilon$=0.11, n=0.49, $\mu_{e}$=16.37, $r_{e}$=0.40\arcsec)
    \item Nuclear disk [\textit{Exponential}] (PA=83.7, $\epsilon$=0.12, $\mu_{e}$=15.65, h=1.56\arcsec)
    \item Nuclear Ring [\textit{GaussianRing2Side}] (PA=86.7, $\epsilon$=0.16, A$_{min}$=19.35,  A$_{maj}$=13.60, $R_{ring}$=3.86\arcsec, $\sigma_{r-in}$=0.27\arcsec, $\sigma_{r-out}$=1.64\arcsec)
    \item B/P Bulge [\textit{Sers\'ic$\_$GenEllipse}] (PA=86.7, $\epsilon$=0.42, n=0.78, $\mu_{e}$=20.19, $r_{e}$=25.13\arcsec)
    \item Bar spurs [\textit{FlatBar}] (PA=90.9, $\epsilon$=0.91, $\mu_{0}$=23.82, R$_{brk}$=80.43\arcsec)
\end{itemize}
\vspace{2pt}
{\bf \large NGC 1440}\\
The NSC in this galaxy is unresolved (see section~\ref{subsec:nucfrac}). For the purpose of determining the NSC properties, we provide a Sers\'ic component fit.
\begin{itemize}
    \item NSC [\textit{Sers\'ic}] (PA=0, $\epsilon$=0.0, n=3.48, $\mu_{e}$=15.34, $r_{e}$=0.05\arcsec)
    \item Classical bulge(?) [\textit{Sers\'ic}] (PA=112.2, $\epsilon$=0.13, n=2.13, $\mu_{e}$=17.28, $r_{e}$=1.73\arcsec)
    \item Nuclear disk [\textit{Sers\'ic}] (PA=106.9, $\epsilon$=0.19, n=64, $\mu_{e}$=18.02, $r_{e}$=1.96\arcsec)
    \item B/P bulge [\textit{Sers\'ic}] (PA=120.0, $\epsilon$=0.16, n=41, $\mu_{e}$=18.51, $r_{e}$=6.34\arcsec)
    \item Bar spurs [\textit{FlatBar}] (PA=140.9, $\epsilon$=0.77, $\mu_{0}$=18.60, R$_{brk}$=18.245\arcsec)
    \item Main disk [\textit{BrokenExponential}] (PA=108.1, $\epsilon$=0.23, $\mu_{0}$=19.93, R$_{brk}$=36.85\arcsec)
\end{itemize}
\vspace{2pt}
{\bf \large NGC 1566}\\
Due to significant nuclear saturation in all bands, we do not provide fits for this galaxy (see section~\ref{subsec:qualitymodel}).\\
\\
{\bf \large NGC 2775}
\begin{itemize}
    \item NSC [\textit{Sers\'ic}] (PA=0, $\epsilon$=0.0, n=0.52, $\mu_{e}$=14.51, $r_{e}$=0.04\arcsec)
    \item Classical bulge(?) [\textit{Sers\'ic}] (PA=23.8, $\epsilon$=0.12, n=2.81, $\mu_{e}$=18.20, $r_{e}$=6.05\arcsec)
    \item Outer bulge [\textit{Sers\'ic}] (PA=26.5, $\epsilon$=0.12, n=1.06, $\mu_{e}$=19.34, $r_{e}$=17.12\arcsec)
    \item Ring [\textit{GaussianRing}] (PA=24.0, $\epsilon$=0.21, A=22.03, $R_{ring}$=31.51\arcsec, $\sigma_{r}$=17.81\arcsec)
    \item Main disk [\textit{Exponential}] (PA=31.7, $\epsilon$=0.22, $\mu_{0}$=19.14, h=42.59\arcsec)
\end{itemize}
\vspace{2pt}
{\bf \large NGC 3351}
\begin{itemize}
    \item NSC [\textit{Sers\'ic}] (PA=0, $\epsilon$=0.0, n=0.8, $\mu_{e}$=15.56, $r_{e}$=0.09\arcsec)
    \item Classical bulge(?) [\textit{Sers\'ic}] (PA=58.1, $\epsilon$=0.32, n=0.42, $\mu_{e}$=16.85, $r_{e}$=0.75\arcsec)
    \item Nuclear disk + ring [\textit{BrokenExponential}] (PA=33.3, $\epsilon$=0.12, $\mu_{0}$=16.88, R$_{brk}$=6\arcsec)
    \item B/P bulge [\textit{Sers\'ic$\_$GenEllipse}] (PA=46.1, $\epsilon$=0.28, n=0.89, $\mu_{e}$=18.89, $r_{e}$=14.58\arcsec)
    \item Bar spurs [\textit{FlatBar}] (PA=142.8, $\epsilon$=0.74, $\mu_{0}$=19.02, R$_{brk}$=41.18\arcsec)
    \item Inner ring [\textit{GaussianRing}] (PA=20.4, $\epsilon$=0.14, A=20.03, $R_{ring}$=59.92\arcsec, $\sigma_{r}$=15.70\arcsec)
    \item Main disk [\textit{BrokenExponential}] (PA=41, $\epsilon$=0.27, $\mu_{0}$=21.25, R$_{brk}$=141.6\arcsec)
\end{itemize}
\vspace{3pt}
{\bf \large NGC 3368}
\begin{itemize}
    \item NSC [\textit{Sers\'ic}] (PA=81.7, $\epsilon$=0.16, n=2.21, $\mu_{e}$=14.29, $r_{e}$=0.204\arcsec)
    \item Classical bulge(?) or inner-bar B/P bulge [\textit{Sers\'ic}] (PA=27.6, $\epsilon$=0.33, n=0.21, $\mu_{e}$=15.86, $r_{e}$=0.93\arcsec)
    \item Inner-bar spurs [\textit{Sers\'ic$\_$GenEllipse}] (PA=141, $\epsilon$=0.61, n=0.31, $\mu_{e}$=16.69, $r_{e}$=3.41\arcsec)
    \item Nuclear disk [\textit{Sers\'ic}] (PA=2.8, $\epsilon$=0.22, n=0.33, $\mu_{e}$=17.13, $r_{e}$=5.42\arcsec)
    \item B/P bulge [\textit{Sers\'ic$\_$GenEllipse}] (PA=-1.8, $\epsilon$=0.28, n=1.73, $\mu_{e}$=18.98, $r_{e}$=30.52\arcsec)
    \item SE outer-bar spur [\textit{Sers\'ic}] (PA=9.7, $\epsilon$=0.54, n=0.70, $\mu_{e}$=21.39, $r_{e}$=37.56\arcsec)
    \item NW outer-bar spur [\textit{Sers\'ic}] (PA=5.6, $\epsilon$=0.48, n=0.73, $\mu_{e}$=22.69, $r_{e}$=32.76\arcsec)
    \item Disk [\textit{BrokenExponential}] (PA=13.9, $\epsilon$=0.36, $\mu_{0}$=20.48, R$_{brk}$=178.65\arcsec)
\end{itemize}
\vspace{3pt}
{\bf \large NGC 3412}
\begin{itemize}
    \item NSC [\textit{Sers\'ic}] (PA=149.0, $\epsilon$=0.31, n=1.78, $\mu_{e}$=13.38, $r_{e}$=0.12\arcsec)
    \item Classical bulge(?) [\textit{Sers\'ic}] (PA=151.0, $\epsilon$=0.36, n=1.05, $\mu_{e}$=15.37, $r_{e}$=0.95\arcsec)
    \item Inner elliptical (counter-rotating) component [\textit{Sers\'ic$\_$GenEllipse}] (PA=148.9, $\epsilon$=0.29, n=0.87, $\mu_{e}$=16.35, $r_{e}$=2.68\arcsec)
    \item B/P bulge [\textit{Sers\'ic$\_$GenEllipse}] (PA=137.2, $\epsilon$=0.22, n=0.56, $\mu_{e}$=18.75, $r_{e}$=5.46\arcsec)
    \item Bar spurs [\textit{Sers\'ic$\_$GenEllipse}] (PA=106.6, $\epsilon$=0.30, n=0.51, $\mu_{e}$=18.63, $r_{e}$=10.79\arcsec)
    \item Disk [\textit{BrokenExponential}] (PA=150.2, $\epsilon$=0.50, $\mu_{0}$=19.81, R$_{brk}$=69.41\arcsec)
    \item Halo [\textit{Sers\'ic}] (PA=144.6, $\epsilon$=0.34, n=0.64, $\mu_{e}$=22.28, $r_{e}$=57.78\arcsec)
\end{itemize}
\vspace{3pt}
{\bf \large NGC 4237}
\begin{itemize}
    \item NSC [\textit{Sers\'ic}] (PA=132.7, $\epsilon$=0.12, n=4.42, $\mu_{e}$=17.56, $r_{e}$=0.21\arcsec)
    \item Bulge [\textit{Sers\'ic}] (PA=121.0, $\epsilon$=0.27, n=0.75, $\mu_{e}$=18.46, $r_{e}$=1.92\arcsec)
   \item Disk [\textit{Exponential}] (PA=121.3, $\epsilon$=0.40, $\mu_{0}$=17.95, h=14.81\arcsec)
\end{itemize}
\vspace{3pt}
{\bf \large NGC 4377}
\begin{itemize}
    \item NSC [\textit{Sers\'ic}] (PA=10.2, $\epsilon$=0.21, n=5.73, $\mu_{e}$=13.60, $r_{e}$=0.02\arcsec)
    \item Classical bulge [\textit{Sers\'ic}] (PA=47.4, $\epsilon$=0.13, n=1.62, $\mu_{e}$=16.10, $r_{e}$=0.74\arcsec)
    \item B/P Bulge [\textit{Sers\'ic$\_$GenEllipse}] (PA=13.1, $\epsilon$=0.16, n=1.27, $\mu_{e}$=16.78, $r_{e}$=2.61\arcsec)
    \item Bar spurs [\textit{FlatBar}] (PA=1.4, $\epsilon$=0.81, $\mu_{0}$=18.80, R$_{brk}$=6.91\arcsec)
    \item Outer ring [\textit{GaussianRing}] (PA=39.8, $\epsilon$=0.14, A =23.12, $R_{ring}$=20.12\arcsec, $\sigma_{r}$=5.70\arcsec)
    \item Disk [\textit{Exponential}] (PA=30.8, $\epsilon$=0.19, $\mu_{0}$=18.20, h=10.89\arcsec)
\end{itemize}
\vspace{3pt}
{\bf \large NGC 4380}
\begin{itemize}
    \item NSC [\textit{Sers\'ic}] (PA=166.7, $\epsilon$=0.18, n=7.0, $\mu_{e}$=16.59, $r_{e}$=0.35\arcsec)
    \item Classical bulge(?) [\textit{Sers\'ic}] (PA=177.1, $\epsilon$=0.21, n=1.73, $\mu_{e}$=17.55, $r_{e}$=2.46\arcsec)
    \item Inner ring [\textit{GaussianRing}] (PA=173.9, $\epsilon$=0.46, A =19.05, $R_{ring}$=12.15\arcsec, $\sigma_{r}$=6.47\arcsec)
    \item Inner disk(?) [\textit{Sers\'ic}] (PA=174.2, $\epsilon$=0.42, n=0.5, $\mu_{e}$=17.63, $r_{e}$=5.50\arcsec)
    \item Outer ring [\textit{GaussianRing}] (PA=169.2, $\epsilon$=0.43, A =20.56, $R_{ring}$=27.42\arcsec, $\sigma_{r}$=3.87\arcsec)
    \item Main disk [\textit{BrokenExponential}] (PA=173.3, $\epsilon$=0.47, $\mu_{0}$=18.60, R$_{brk}$=71.48\arcsec)
\end{itemize}
\vspace{3pt}
{\bf \large NGC 4450}
\begin{itemize}
    \item NSC [\textit{Sers\'ic}] (PA=0, $\epsilon$=0.0, n=4.95, $\mu_{e}$=15.73, $r_{e}$=0.19\arcsec)
    \item Classical bulge(?) or inner-bar B/P bulge [\textit{Sers\'ic}] (PA=86.7, $\epsilon$=0.21, n=1.21, $\mu_{e}$=16.06, $r_{e}$=0.64\arcsec)
    \item Inner bar (spurs?) [\textit{Sers\'ic}] (PA=46.5, $\epsilon$=0.12, n=0.89, $\mu_{e}$=17.07, $r_{e}$=1.62\arcsec)
    \item Nuclear ring [\textit{GaussianRing}] (PA=168.6, $\epsilon$=0.43, A =18.99, $R_{ring}$=3.54\arcsec, $\sigma_{r}$=1.01\arcsec)
    \item Nuclear disk [\textit{Exponential}] (PA=81.8, $\epsilon$=0.22, $\mu_{0}$=16.34, h=3.24\arcsec)
    \item Outer-bar B/P bulge [\textit{Sers\'ic$\_$GenEllipse}] (PA=87.1, $\epsilon$=0.35, n=1.0, $\mu_{e}$=19.53, $r_{e}$=28.30\arcsec)
    \item N outer-bar spur [\textit{Sers\'ic$\_$GenEllipse}] (PA=49.4, $\epsilon$=0.37, n=0.85, $\mu_{e}$=22.71, $r_{e}$=22.34\arcsec)
    \item S outer-bar spur [\textit{Sers\'ic$\_$GenEllipse}] (PA=51.7, $\epsilon$=0.38, n=1.0, $\mu_{e}$=22.43, $r_{e}$=31.0\arcsec)
    \item Disk [\textit{BrokenExponential}] (PA=166.0, $\epsilon$=0.33, $\mu_{0}$=21.11, R$_{brk}$=108.27\arcsec)
\end{itemize}
\vspace{3pt}
{\bf \large NGC 4501}
\begin{itemize}
    \item NSC [\textit{Sers\'ic}] (PA=144.2, $\epsilon$=0.22, n=1.58, $\mu_{e}$=13.33, $r_{e}$=0.38\arcsec)
    \item Bulge [\textit{Sers\'ic}] (PA=135.9, $\epsilon$=0.33, n=3.02, $\mu_{e}$=16.93, $r_{e}$=10.67\arcsec)
   \item Disk [\textit{Exponential}] (PA=140.4, $\epsilon$=0.51, $\mu_{0}$=16.24, h=41.17\arcsec)
\end{itemize}
\vspace{3pt}
{\bf \large NGC 4531}
\begin{itemize}
    \item NSC [\textit{Sers\'ic}] (PA=39.5, $\epsilon$=0.05, n=4.36, $\mu_{e}$=16.59, $r_{e}$=0.15\arcsec)
    \item Unclear [\textit{Exponential}] (PA=71.1, $\epsilon$=0.73, $\mu_{0}$=21.29, h=10.18\arcsec)
    \item Inner disk/pseudobulge(?) [\textit{Sers\'ic}] (PA=152.1, $\epsilon$=0.23, n=1.24, $\mu_{e}$=21.06, $r_{e}$=15.06\arcsec)
    \item Ring [\textit{GaussianRing}] (PA=24.3, $\epsilon$=0.22, A =21.92, $R_{ring}$=17.98\arcsec, $\sigma_{r}$=2.83\arcsec)
    \item Disk [\textit{Exponential}] (PA=14.8, $\epsilon$=0.46, $\mu_{0}$=19.27, h=29.62\arcsec)
\end{itemize}
\vspace{3pt}
{\bf \large NGC 4548}
\begin{itemize}
    \item NSC [\textit{Sers\'ic}] (PA=0, $\epsilon$=0.20, n=2.47, $\mu_{e}$=14.35, $r_{e}$=0.25\arcsec)
    \item Inner-bar B/P bulge(?) [\textit{Sers\'ic}] (PA=157.1, $\epsilon$=0.03, n=0.60, $\mu_{e}$=15.86, $r_{e}$=0.62\arcsec)
    \item Inner bar [\textit{Sers\'ic}] (PA=132.7, $\epsilon$=0.30, n=0.50, $\mu_{e}$=17.60, $r_{e}$=2.01\arcsec)
    \item Nuclear disk [\textit{Sers\'ic}] (PA=158.2, $\epsilon$=0.12, n=1.0, $\mu_{e}$=18.19, $r_{e}$=5.66\arcsec)
    \item Outer-bar B/P bulge [\textit{Sers\'ic}] (PA=113.5, $\epsilon$=0.15, n=0.37, $\mu_{e}$=19.76, $r_{e}$=16.90\arcsec)
    \item Outer-bar spurs [\textit{FlatBar}] (PA=97.7, $\epsilon$=0.80, $\mu_{0}$=20.42, R$_{brk}$=41.16\arcsec)
    \item W spiral arm [\textit{LogSpiralBrokenExp}] (PA=0, $\epsilon$=0, $R_{i}$=124.38\arcsec, $\sigma_{az}$=17.81\arcsec, $\mu_{max}$=19.78, $r_{b}$=108.97\arcsec, $R_{max}$=70.13\arcsec, $\sigma_{trunc}$=56.99\arcsec)
    \item E spiral arm [\textit{LogSpiralBrokenExp}] (PA=163.0, $\epsilon$=0.14, $R_{i}$=115.95\arcsec, $\sigma_{az}$=29.31\arcsec, $\mu_{max}$=19.67, $r_{b}$=134.24\arcsec, $R_{max}$=70.13\arcsec, $\sigma_{trunc}$=14.32\arcsec)
    \item Disk [\textit{BrokenExponential}] (PA=173.0, $\epsilon$=0.21, $\mu_{0}$=20.33, R$_{brk}$=116.09\arcsec)
\end{itemize}
\vspace{3pt}
{\bf \large NGC 4578}
\begin{itemize}
    \item NSC [\textit{Sers\'ic}] (PA=42.0, $\epsilon$=0.14, n=2.76, $\mu_{e}$=14.89, $r_{e}$=0.15\arcsec)
    \item Classical bulge [\textit{Sers\'ic}] (PA=64.8, $\epsilon$=0.23, n=2.47, $\mu_{e}$=18.39, $r_{e}$=4.96\arcsec)
    \item Nuclear ring [\textit{GaussianRing}] (PA=62.3, $\epsilon$=0.27, A =18.50, $R_{ring}$=0.75\arcsec, $\sigma_{r}$=3.74\arcsec)
    \item Outer ring [\textit{GaussianRing}] (PA=64.9, $\epsilon$=0.31, A =21.88, $R_{ring}$=44.99\arcsec, $\sigma_{r}$=27.51\arcsec)
    \item Disk [\textit{Exponential}] (PA=62.5, $\epsilon$=0.32, $\mu_{0}$=18.46, h=12.92\arcsec)
\end{itemize}
\vspace{3pt}
{\bf \large NGC 4579}
\begin{itemize}
    \item NSC [\textit{Sers\'ic}] (PA=95.0, $\epsilon$=0.30, n=1.54, $\mu_{e}$=10.52, $r_{e}$=0.07\arcsec)
    \item Classical bulge or nuclear disk [\textit{Sers\'ic}] (PA=118.2, $\epsilon$=0.19, n=2.49, $\mu_{e}$=15.53, $r_{e}$=5.18\arcsec)
    \item B/P Bulge [\textit{Sers\'ic}] (PA=93.7, $\epsilon$=0.36, n=0.78, $\mu_{e}$=18.53, $r_{e}$=20.78\arcsec)
    \item Bar spurs [\textit{FlatBar}] (PA=86.3, $\epsilon$=0.90, $\mu_{0}$=17.33, R$_{brk}$=30.13\arcsec)
    \item Disk [\textit{Exponential}] (PA=125.0, $\epsilon$=0.20, $\mu_{0}$=17.42, h=47.82\arcsec)
\end{itemize}
\vspace{3pt}
{\bf \large NGC 4608}
\begin{itemize}
    \item NSC [\textit{Sers\'ic}] (PA=164.0, $\epsilon$=0.13, n=1.19, $\mu_{e}$=13.80, $r_{e}$=0.03\arcsec)
    \item Classical bulge [\textit{Sers\'ic}] (PA=105.7, $\epsilon$=0.05, n=2.75, $\mu_{e}$=18.96, $r_{e}$=8.01\arcsec)
    \item B/P bulge [\textit{Sers\'ic$\_$GenEllipse}] (PA=46.7, $\epsilon$=0.18, n=1.07, $\mu_{e}$=19.42, $r_{e}$=11.40258\arcsec)
    \item Bar spurs [\textit{FlatBar}] (PA=44.9, $\epsilon$=0.87, $\mu_{0}$=19.40, R$_{brk}$=40.71\arcsec)
    \item Inner ring [\textit{GaussianRingAz}] (PA=124.5, $\epsilon$=0.09, A$_{maj}$ =20.46, A$_{min-rel}$=22.07,  $R_{ring}$=48.03\arcsec, $\sigma_{r}$=6.44\arcsec)
    \item Disk [\textit{BrokenExponential}] (PA=124.9, $\epsilon$=0.13, $\mu_{0}$=23.82, R$_{brk}$=69.78\arcsec)
\end{itemize}
\vspace{3pt}
{\bf \large NGC 4612}\\
The NSC in this galaxy is fitted using a two-component model. The properties of this NSC provided in Table~\ref{tab:nscprops} are integrated over the two NSC components below.
\begin{itemize}
    \item Inner NSC [\textit{Sers\'ic}] (PA=0, $\epsilon$=0.0, n=3.14, $\mu_{e}$=12.39, $r_{e}$=0.03\arcsec)
    \item Outer NSC [\textit{Exponential}] (PA=23.25, $\epsilon$=0.22, $\mu_{0}$=14.05, h=0.14\arcsec)
    \item Nuclear disk [\textit{Exponential}] (PA=26.9, $\epsilon$=0.22, $\mu_{0}$=14.84, h=1.16\arcsec)
    \item B/P bulge [\textit{Exponential$\_$GenEllipse}] (PA=33.3, $\epsilon$=0.30, $\mu_{0}$=17.90, h=5.01\arcsec)
    \item Bar spurs [\textit{FlatBar}] (PA=156.7, $\epsilon$=0.35, $\mu_{0}$=18.37, R$_{brk}$=16.23\arcsec)
    \item Inner ring [\textit{GaussianRing}] (PA=21.9, $\epsilon$=0.40, A =21.69, $R_{ring}$=28.02\arcsec, $\sigma_{r}$=8.13\arcsec)
    \item Disk [\textit{BrokenExponential}] (PA=15.7, $\epsilon$=0.27, $\mu_{0}$=20.02, R$_{brk}$=48.51\arcsec)
\end{itemize}
\vspace{3pt}
{\bf \large NGC 4643}
\begin{itemize}
    \item NSC [\textit{Sers\'ic}] (PA=0, $\epsilon$=0.0, n=0.76, $\mu_{e}$=14.37, $r_{e}$=0.04\arcsec)
    \item Classical bulge(?) [\textit{Sers\'ic}] (PA=96.6, $\epsilon$=0.11, n=0.69, $\mu_{e}$=15.79, $r_{e}$=0.32\arcsec)
    \item Nuclear disk [\textit{BrokenExponential}] (PA=118.3, $\epsilon$=0.13, $\mu_{0}$=15.19, R$_{brk}$=2.99\arcsec)
    \item B/P bulge [\textit{Sers\'ic}] (PA=8.4, $\epsilon$=0.13, n=0.62, $\mu_{e}$=18.44, $r_{e}$=12.92\arcsec)
    \item Bar spurs [\textit{FlatBar}] (PA=18.9, $\epsilon$=0.90, $\mu_{0}$=19.08, R$_{brk}$=44.49\arcsec)
    \item Disk [\textit{Exponential}] (PA=124.6, $\epsilon$=0.18, $\mu_{0}$=20.55, h=81.27\arcsec)
\end{itemize}
\vspace{3pt}
{\bf \large NGC 4689}\\
The precise structural/morphological nature of the 2nd--4th components in our model is currently unclear, so we only refer to them as ``inner components''.
\begin{itemize}
    \item NSC [\textit{Sers\'ic}] (PA=0, $\epsilon$=0.0, n=2.85, $\mu_{e}$=15.33, $r_{e}$=0.06\arcsec)
    \item Inner component 1 [\textit{Sers\'ic}] (PA=13.4, $\epsilon$=0.22, n=0.70, $\mu_{e}$=19.91, $r_{e}$=0.91\arcsec)
    \item Inner component 2 [\textit{Sers\'ic}] (PA=2.8, $\epsilon$=0.14, n=1.47, $\mu_{e}$=20.63, $r_{e}$=6.74\arcsec)
    \item Inner component 3 [\textit{Sers\'ic}] (PA=79.6, $\epsilon$=0.13, n=0.88, $\mu_{e}$=20.63, $r_{e}$=7.19\arcsec)
    \item Ring [\textit{GaussianRing}] (PA=66.5, $\epsilon$=0.30, A =21.81, $R_{ring}$=16.92\arcsec, $\sigma_{r}$=2.97\arcsec)
    \item Disk [\textit{Exponential}] (PA=87.7, $\epsilon$=0.26, $\mu_{0}$=19.75, h=43.68\arcsec)
\end{itemize}
\vspace{3pt}
{\bf \large NGC 4698}
\begin{itemize}
    \item NSC [\textit{Sers\'ic}] (PA=76.6, $\epsilon$=0.48, n=1.39, $\mu_{e}$=15.61, $r_{e}$=0.10\arcsec)
    \item Orthogonal bulge [\textit{Sers\'ic}] (PA=80.2, $\epsilon$=0.53, n=1.96, $\mu_{e}$=17.64, $r_{e}$=2.36\arcsec)
    \item Classical bulge(?) [\textit{Sers\'ic}] (PA=171.3, $\epsilon$=0.08, n=3.10, $\mu_{e}$=19.78, $r_{e}$=20.01\arcsec)
    \item Inner ring [\textit{GaussianRing2Side}] (PA=176.9, $\epsilon$=0.67, A=21.13,  $R_{ring}$=40.39\arcsec, $\sigma_{r-in}$=8.62\arcsec, $\sigma_{r-out}$=15.29\arcsec)
    \item Outer ring [\textit{GaussianRing}] (PA=178.9, $\epsilon$=0.67, A =22.86, $R_{ring}$=64.25\arcsec, $\sigma_{r}$=3.41\arcsec)
    \item Disk [\textit{BrokenExponential}] (PA=176.5, $\epsilon$=0.63, $\mu_{0}$=22.10, R$_{brk}$=67.91\arcsec)
    \item Halo [\textit{Sers\'ic}] (PA=179.3, $\epsilon$=0.32, n=1.21, $\mu_{e}$=21.99, $r_{e}$=54.79\arcsec)
\end{itemize}
\vspace{3pt}
{\bf \large NGC 4699}
\begin{itemize}
    \item NSC [\textit{Sers\'ic}] (PA=0, $\epsilon$=0.0, n=1.87, $\mu_{e}$=14.62, $r_{e}$=0.28\arcsec)
    \item Classical bulge(?) [\textit{Sers\'ic}] (PA=73.3, $\epsilon$=0.21, n=1.13, $\mu_{e}$=14.79, $r_{e}$=1.52\arcsec)
    \item B/P bulge [\textit{Sers\'ic$\_$GenEllipse}] (PA=73.2, $\epsilon$=0.39, n=0.43, $\mu_{e}$=16.48, $r_{e}$=3.21\arcsec)
    \item Bar spurs [\textit{Sers\'ic$\_$GenEllipse}] (PA=78.4, $\epsilon$=0.56, n=0.26, $\mu_{e}$=17.49, $r_{e}$=7.93\arcsec)
    \item NE bar ansa [\textit{Sers\'ic}] (PA=8.1, $\epsilon$=0.38, n=0.88, $\mu_{e}$=19.13, $r_{e}$=3.60\arcsec)
    \item SW bar ansa [\textit{Sers\'ic}] (PA=87.7, $\epsilon$=0.35, n=0.96, $\mu_{e}$=19.40, $r_{e}$4.18\arcsec)
    \item Inner disk [\textit{Exponential}] (PA=67.5, $\epsilon$=0.26, $\mu_{0}$=16.29, h=12.58\arcsec)
    \item Ring [\textit{GaussianRing2Side}] (PA=67.7, $\epsilon$=0.38, A=21.17,  $R_{ring}$=49.39\arcsec, $\sigma_{r-in}$=2.38\arcsec, $\sigma_{r-out}$=18.04\arcsec)
    \item Main disk [\textit{Exponential}] (PA=57.8, $\epsilon$=0.14, $\mu_{0}$=20.14, h=67.34\arcsec)
\end{itemize}
\vspace{3pt}
{\bf \large NGC 5121}
\begin{itemize}
    \item NSC [\textit{Sers\'ic}] (PA=35.2, $\epsilon$=0.14, n=1.26, $\mu_{e}$=13.84, $r_{e}$=0.07\arcsec)
    \item Classical bulge(?) or nuclear-bar B/P bulge [\textit{Sers\'ic}] (PA=111.1, $\epsilon$=0.17, n=0.57, $\mu_{e}$=15.32, $r_{e}$=0.28\arcsec)
    \item Nuclear bar [\textit{FlatBar}] (PA=45.5, $\epsilon$=0.74, $\mu_{0}$=15.57, R$_{brk}$=1.13\arcsec)
    \item Nuclear disk [\textit{Exponential}] (PA=57.1, $\epsilon$=0.12, $\mu_{0}$=14.27, h=0.87\arcsec)
    \item Outer-bar B/P bulge(?) [\textit{Sers\'ic$\_$GenEllipse}] (PA=84.2, $\epsilon$=0.03, n=0.44, $\mu_{e}$=18.18, $r_{e}$=3.90\arcsec)
    \item Outer-bar spurs(?) [\textit{Sers\'ic}] (PA=51.3, $\epsilon$=0.25, n=0.21, $\mu_{e}$=19.69, $r_{e}$=8.98\arcsec)
    \item Disk [\textit{BrokenExponential}] (PA=57.0, $\epsilon$=0.21, $\mu_{0}$=18.45, R$_{brk}$=36.23\arcsec)
\end{itemize}
\vspace{3pt}
{\bf \large NGC 5248}
\begin{itemize}
    \item NSC [\textit{Sers\'ic}] (PA=128.3, $\epsilon$=0.27, n=0.97, $\mu_{e}$=15.60, $r_{e}$=0.41\arcsec)
    \item Classical bulge(?) [\textit{Sers\'ic}] (PA=156.2, $\epsilon$=0.30, n=0.38, $\mu_{e}$=17.71, $r_{e}$=1.02\arcsec)
    \item Nuclear disk [\textit{Exponential}] (PA=125.5, $\epsilon$=0.32, $\mu_{0}$=16.34, h=4.34\arcsec)
    \item Boxy zone [\textit{Sers\'ic$\_$GenEllipse}] (PA=123.0, $\epsilon$=0.44, n=0.21, $\mu_{e}$=20.20, $r_{e}$=18.75\arcsec)
    \item N inner spiral [\textit{LogSpiralArc}] (PA=0, $\epsilon$=0, r$_scale$=64.17, $\mu_{max}$=20.39, $\sigma_{r}$=10.60\arcsec, $\sigma_{\theta_ccw}$=0.45\arcsec, $\sigma_{\theta_cw}$=9.0\arcsec)
    \item S inner spiral [\textit{LogSpiralArc}] (PA=0 $\epsilon$=0, r$_scale$=52.55, $\mu_{max}$=20.62, $\sigma_{r}$=11.23\arcsec, $\sigma_{\theta_ccw}$=1.96\arcsec, $\sigma_{\theta_cw}$=3.19\arcsec)
    \item Bar [\textit{Sers\'ic}] (PA=157.4, $\epsilon$=0.36, n=1.0, $\mu_{e}$=21.44, $r_{e}$=66.78\arcsec)
\end{itemize}
\vspace{3pt}
{\bf \large NGC 5364}
\begin{itemize}
    \item NSC [\textit{Sers\'ic}] (PA=0, $\epsilon$=0.16, n=3.52, $\mu_{e}$=17.36, $r_{e}$=0.21\arcsec)
    \item Classical bulge(?) [\textit{Sers\'ic}] (PA=55.4, $\epsilon$=0.26, n=0.65, $\mu_{e}$=18.66, $r_{e}$=0.69\arcsec)
    \item Pseudobulge(?) [\textit{Sers\'ic}] (PA=28.3, $\epsilon$=0.17, n=1.0, $\mu_{e}$=19.90, $r_{e}$=6.26\arcsec)
    \item Ring [\textit{GaussianRing}] (PA=70.1, $\epsilon$=0.49, A=21.84, $R_{ring}$=31.04\arcsec, $\sigma_{r}$=9.39\arcsec)
    \item Inner disk [\textit{Exponential}] (PA=67.8, $\epsilon$=0.46, $\mu_{0}$=19.56, h=25.80\arcsec)
    \item Outer disk [\textit{Exponential}] (PA=28.6, $\epsilon$=0.32, $\mu_{0}$=20.49, h=59.47\arcsec)
\end{itemize}
\vspace{3pt}
{\bf \large NGC 6744}
\begin{itemize}
    \item NSC [\textit{Sers\'ic}] (PA=41.7, $\epsilon$=0.38, n=2.28, $\mu_{e}$=14.12, $r_{e}$=0.25\arcsec)
    \item Classical bulge [\textit{Sers\'ic}] (PA=38.4, $\epsilon$=0.15, n=3.05, $\mu_{e}$=17.35, $r_{e}$=7.18\arcsec)
    \item B/P bulge [\textit{Sers\'ic$\_$GenEllipse}] (PA=31.6, $\epsilon$=0.38, n=0.89, $\mu_{e}$=17.41, $r_{e}$=15.52\arcsec)
    \item Bar spurs [\textit{Sers\'ic}] (PA=25.6, $\epsilon$=0.74, n=0.64, $\mu_{e}$=19.33, $r_{e}$=71.10\arcsec)
    \item Disk [\textit{BrokenExponential}] (PA=39.5, $\epsilon$=0.31, $\mu_{0}$=19.02, R$_{brk}$=99.73\arcsec)
\end{itemize}
\vspace{3pt}
{\bf \large NGC 7177}\\
We do not trust the NSC measurements for this galaxy due to strong dust lanes obscuring the NSC, as mentioned in Section~\ref{subsec:qualitymodel}. Hence, we do not provide the best-fit models for this galaxy.\\\\
{\bf \large NGC 7513}
\begin{itemize}
    \item NSC [\textit{Sers\'ic}] (PA=150.0, $\epsilon$=0.0, n=5.92, $\mu_{e}$=19.29, $r_{e}$=0.45\arcsec)
    \item Classical bulge(?) [\textit{Sers\'ic}] (PA=133.3, $\epsilon$=0.18, n=0.66, $\mu_{e}$=19.98, $r_{e}$=1.97\arcsec)
    \item B/P bulge [\textit{Sers\'ic$\_$GenEllipse}] (PA=151.4, $\epsilon$=0.26, n=0.96, $\mu_{e}$=20.01, $r_{e}$=8.58\arcsec)
    \item Bar spurs [\textit{FlatBar}] (PA=145.5, $\epsilon$=0.92, $\mu_{0}$=19.39, R$_{brk}$=30.31\arcsec)
    \item Disk [\textit{BrokenExponential}] (PA=3.4, $\epsilon$=0.33, $\mu_{0}$=21.08, R$_{brk}$=62.04\arcsec)
\end{itemize}
\vspace{3pt}
\end{document}